\documentclass[aps, prd, twocolumn, superscriptaddress, showpacs, floatfix]{revtex4}
 
\usepackage{epsf}
\usepackage{bm}
\usepackage{amssymb}
\usepackage{amsmath}
\usepackage{amsfonts}
\usepackage{dcolumn}
\usepackage{natbib}
\usepackage{color}
\usepackage[dvips]{graphicx}

\newcommand{\der}[2]{\frac{\partial #1}{\partial #2}}

\newcommand{\w}[1]{\bm{#1}}
\newcommand{\M}{{\mathcal M}}
\newcommand{\Sp}{{\mathcal S}}

\newcommand{\Lie}[1]{\bm{\mathcal L}_{\w{#1}}\,}

\newcommand{\DSc}{\mathcal{D}}

\newenvironment{equationarray}
{\arraycolsep 0.14 em
\begin{eqnarray}}
{\end{eqnarray}}

\newenvironment{equationarray*}
{\arraycolsep 0.14 em
\begin{eqnarray*}}
{\end{eqnarray*}}

\begin{document}

\title{Improved constrained scheme for the Einstein equations:
  An approach to the uniqueness issue}

\author{Isabel Cordero-Carri\'on}
\email{Isabel.Cordero@uv.es}
\affiliation{Departamento de Astronom\'\i a y Astrof\'\i sica,
  Universidad de Valencia, C/ Dr.~Moliner 50, E-46100 Burjassot,
  Valencia, Spain}

\author{Pablo Cerd\'a-Dur\'an}
\email{cerda@mpa-garching.mpg.de}
\affiliation{Max-Planck-Institut f\"ur Astrophysik,
  Karl-Schwarzschild-Strasse~1, D-85741 Garching, Germany}

\author{Harald Dimmelmeier}
\email{harrydee@mpa-garching.mpg.de}
\affiliation{Department of Physics, Aristotle University of
  Thessaloniki, GR-54124 Thessaloniki, Greece}

\author{Jos\'e Luis Jaramillo}
\email{jarama@iaa.es}
\affiliation{Instituto de Astrof\'{\i}sica de Andaluc\'{\i}a, CSIC,
  Apartado Postal 3004, E-18080 Granada, Spain}
\affiliation{Laboratoire Univers et Th\'eories, Observatoire de
  Paris, CNRS, Universit\'e Paris Diderot, 5 place Jules Janssen,
  F-92190 Meudon, France}

\author{J\'er\^ome Novak}
\email{Jerome.Novak@obspm.fr}
\affiliation{Laboratoire Univers et Th\'eories, Observatoire de
  Paris, CNRS, Universit\'e Paris Diderot, 5 place Jules Janssen,
  F-92190 Meudon, France}

\author{Eric Gourgoulhon}
\email{eric.gourgoulhon@obspm.fr}
\affiliation{Laboratoire Univers et Th\'eories, Observatoire de
  Paris, CNRS, Universit\'e Paris Diderot, 5 place Jules Janssen,
  F-92190 Meudon, France}

\date{15 December 2008}


\begin{abstract}
  Uniqueness problems in the elliptic sector of constrained
  formulations of Einstein equations have a dramatic effect on the
  physical validity of some numerical solutions, for instance, when
  calculating the spacetime of very compact stars or nascent black
  holes. The fully constrained formulation (FCF) proposed by Bonazzola,
  Gourgoulhon, Grandcl\'ement, and Novak is one
  of these formulations. It contains, as a particular case, the
  approximation of the conformal flatness condition (CFC) which, in
  the last ten years, has been used in many astrophysical
  applications. The elliptic part of the FCF basically shares the same
  differential operators as the elliptic equations in CFC scheme.
  We present here a reformulation of the elliptic sector of
  CFC that has the fundamental property of overcoming the local
  uniqueness problems. The correct behavior of our new formulation is
  confirmed by means of a battery of numerical simulations. Finally,
  we extend these ideas to FCF, complementing the
  mathematical analysis carried out in previous studies.
\end{abstract}


\maketitle


\section{Introduction}
\label{section:introduction}

In recent years we have seen the successful application of numerical
codes to accurately calculate the spacetimes of compact astrophysical
objects like collapsing stellar cores, (proto)neutron stars, and
black holes. Most of these codes are based on the
$ 3 + 1 $ formalism of general relativity (see, e.g., \cite{York79,
Alcub08, Gourgoulhon07a} for reviews). They typically fall into two
classes. One approach relies on the \emph{free evolution} of the
$ 3 + 1 $ Einstein
equations, recast in order to cure long-term stability problems. 
Here the constraint equations are only solved initially, and
closely monitored at each time step to control the accuracy of the numerical
solution.

Alternatively, formulations based on a \emph{constrained evolution}, where the
constraints are solved in parallel with evolution equations, 
have proven to be successful as well. Such
approaches exhibit the advantage that the solution cannot violate the
constraints by definition (within the accuracy of the numerical
scheme). In particular, the conformally flat
approximation~\cite{Isenberg, WilsoM89} (hereafter CFC) of the full
Einstein equations, which constitutes a fully constrained formulation,
has been shown to yield long-term stable evolutions of such astrophysical
scenarios (see, e.g., \cite{Dimmelmeier02b, oechslin_02_a, saijo_04_a,
  abdikamalov_08_a}). However, apart from computational challenges, arising
from the need to frequently solve the elliptic constraint equations,
constrained formulations suffer from mathematical nonuniqueness
problems when the configuration becomes too compact. In the case of
the collapse of a stellar core or a (proto)neutron star to a black
hole, such a situation is encountered already before the apparent
horizon forms. This issue has, in the past, been prohibitive to
successfully applying such formulations in numerical simulations of a
wide range of astrophysical problems.

The nonuniqueness of solutions stems from the non-linearity of the constraint equations
and has been studied within the so-called \emph{extended conformal
  thin sandwich (XCTS)}~\cite{Yor99, PfeYor03, Pfe05} approach to the
initial data problem in general relativity. In Ref.~\cite{Pfeiffer05}
a parabolic branching was
numerically found in the solutions to the XCTS equations for perturbations of
Minkowski spacetime, providing the first evidence of nonuniqueness in this
elliptic system. First analytical studies have been carried out
in~\cite{BauOMuPfe07, Wal07}, finding support for the genericity of this
nonuniqueness behavior. More specifically, the XCTS elliptic system is
formed by the Einstein constraint equations in a \emph{conformal thin sandwich (CTS)}
decomposition~\cite{Yor99} supplemented with an additional elliptic equation
for the lapse function, which follows from the maximal slicing condition.
Although no general results on existence and uniqueness for the XCTS system
are available (in contrast to the CTS case and similar elliptic
systems encompassing only the constraints; see, e.g., \cite{OMuYor73, OMuYor74,
  York79, Yor99, PfeYor03}), the analysis in~\cite{BauOMuPfe07}
strongly suggests the presence of a \emph{wrong} sign in a
certain term of the lapse equation as the culprit for the loss of
uniqueness, essentially because it spoils the application of a maximum
principle to guarantee uniqueness. Moreover, in these circumstances
(namely, the existence of a nontrivial kernel for the XCTS elliptic operator)
it is shown in~\cite{Wal07} that the parabolic behavior found
in~\cite{Pfeiffer05} is indeed generic.

Certain constrained evolution formalisms which incorporate elliptic gauges in
their schemes
contain elliptic subsystems which share essential points with the XCTS
equations. Nonuniqueness in the elliptic subsystem is certainly an issue for
the well-posedness of the whole elliptic-hyperbolic evolution system. In
numerical implementations this can depend on the employed numerical scheme,
in particular, on its capability to remain \emph{close} to one of the solutions,
at least as long as the solution stays sufficiently far from the branching
point. In fact, constrained or partially constrained evolutions have shown to
be robust in a variety of contexts (see, e.g., the references in~\cite{Rinne08} and
Sec.~5.2.2 of~\cite{JarValGou08}). However, the problems described above
have also emerged, for instance, in the axisymmetric case
in~\cite{ChoHirLie03b, Rinne06} (see also~\cite{RinSte05}). The
analysis in~\cite{Rinne08} concludes that the reason behind the failures in
these axisymmetric formulations is in fact related to the presence of \emph{wrong}
signs or, more precisely, to the \emph{indefinite} character of certain
non-linear Helmholtz-like equations present in the scheme (see~\cite{Rinne08}
for details and also for a parallel numerical discussion in terms of a class
of relaxation methods for the convergence of the elliptic solvers). Regarding
the full three-dimensional case, fully constrained formalisms have been
presented in~\cite{Bonazzola04, Moncr07, Cordero08}. While the work
in~\cite{Bonazzola04, Cordero08} includes an elliptic subsystem closely
related to the XCTS equations and therefore suffers potentially from these
nonuniqueness problems, the uniqueness properties of the scheme of~\cite{Moncr07}
must yet be studied. In both cases, the full numerical
performance still has to be assessed.

The goal of the present work is to discuss a scheme addressing the nonuniqueness
issues of XCTS-like elliptic systems in the full three-dimensional case, with
astrophysical applications as our main motivation. Having the analysis of the
fully constrained formalism (hereafter \emph{FCF}) of~\cite{Bonazzola04, Cordero08} as our
ultimate aim, we focus on an approximation in the
spirit of the CFC approximation by Isenberg, Wilson,
and Mathews~\cite{Isenberg, Wilson96}. This methodological choice is justified since the
CFC scheme already contains the relevant elliptic system of FCF, but in a
setting in which potential additional problematic issues related to the FCF hyperbolic part 
do not mix up with the specific problem we are addressing here.
Therefore, we discuss in detail a modification of the CFC scheme (in the
presence of matter) where maximum-principle lines of reasoning can be used to
infer the uniqueness of the solutions. We investigate numerically the
performance of the new CFC scheme and finally indicate the main lines for its
generalization to the full Einstein FCF case.

The article is organized as follows. In Sec.~\ref{ss:FCF_CFC} we
review the FCF and CFC formalisms, and then discuss the limitations found in the
numerical implementations of the latter. In Sec.~\ref{sect:schemeCFC}
we introduce the modification of the CFC scheme, with the aim of solving the uniqueness
issues, and we present various numerical tests of the new scheme in Sec.~\ref{s:numerical_results}. 
In Sec.~\ref{ss:Gen_FCF} the guidelines
for the generalization to the FCF case are discussed, and conclusions
are drawn in Sec.~\ref{s:discussion}. In the
Appendix we justify a further approximation assumed
in Sec.~\ref{sect:schemeCFC} which is consistent with the CFC setting.
Throughout the paper we use the signature $ (-, +, +, +) $ for the spacetime metric, and
units in which $ c = G = M_\odot = 1 $. Greek indices run from 0 to 3, whereas Latin ones
from 1 to 3 only.


\section{The fully constrained formalism and the conformal flatness condition}
\label{ss:FCF_CFC}


\subsection{A brief review of the fully constrained formalism}
\label{ss:brief_review}

Given an asymptotically flat spacetime $(\M, g_{\mu\nu})$ we consider
a $ 3 + 1 $ splitting by spacelike hypersurfaces $\Sigma_t$, 
denoting timelike unit normals to $\Sigma_t$ by $n^\mu$.
The data on each spacelike hypersurface  $\Sigma_t$ are given by the pair 
$(\gamma_{ij}, K^{ij})$, where $\gamma_{\mu\nu}= g_{\mu\nu}+n_\mu n_\nu$ 
is the Riemannian metric induced on $\Sigma_t$. We choose the convention 
$K_{\mu\nu}=-\frac{1}{2}{\cal L}_{\w n}\gamma_{\mu\nu}$ for the extrinsic curvature.
With the lapse function $N$ and the shift vector $\beta^i$,
the Lorentzian metric $g_{\mu\nu}$ can be expressed in coordinates
$(x^\mu)$ as
\begin{equation}
  g_{\mu\nu} \, dx^\mu \, dx^\nu =
  - N^2 \, dt^2 + \gamma_{ij} (dx^i + \beta^i \, dt)(dx^j + \beta^j \, dt).
\end{equation}
On the other hand, we can write
\begin{equation}
  2 N K^{ij} = \partial_t \gamma^{ij} + D^i \beta^j + D^j \beta^i,
\end{equation}
where  $D_i$ is the Levi--Civita connection associated with $\gamma_{\mu\nu}$
and  $\partial_t \gamma^{ij}$ represents the Lie derivative
with respect to the evolution vector $t^\mu:= (\partial_t)^\mu=N n^\mu+\beta^\mu$.
As in~\cite{Bonazzola04} we introduce a \emph{time independent} flat metric
$f_{ij}$, which
satisfies ${\cal L}_{\w{t}}f_{ij}=\partial_t f_{ij}=0$ and
coincides with $\gamma_{ij}$ at spatial infinity. We define
$\gamma := \det \gamma_{ij}$ and $f := \det f_{ij}$. This fiducial metric 
permits the use of tensor quantities rather than tensor densities. The
next step in the formulation of~\cite{Bonazzola04} is the conformal decomposition of
the $ 3 + 1 $ fields. First, a representative $\tilde{\gamma}_{ij}$ 
in the conformal class of $\gamma_{ij}$ is chosen, so we can write
\begin{equationarray}
  \gamma_{ij}  = 
  \psi^4 \tilde{\gamma}_{ij},
  \qquad
  K^{ij} = \psi^{\zeta-8} \tilde{A}^{ij} + \frac{1}{3} K \gamma^{ij},
  \label{e:conformal_decomp}
\end{equationarray}%
where $K=\gamma^{ij}K_{ij}$ and $\tilde{\gamma}:=\det
\tilde{\gamma}_{ij}$, and $\zeta\in\mathbb{R}$. 
In Ref.~\cite{Bonazzola04}, the choice $\zeta=4$ was adopted, leading to 
the following expression 
of $\tilde{A}^{ij}$ in terms of the lapse $N$ and shift $\beta^i$: 
\begin{equation}
  \tilde{A}^{ij} = \frac{1}{2 N}
  \left( \tilde{D}^i \beta^j + \tilde{D}^j \beta^i -
  \frac{2}{3} \tilde{D}_k \beta^k \tilde{\gamma}^{ij} +
  \partial_t \tilde{\gamma}^{ij} \right),
  \label{e:Aij}
\end{equation}
$\tilde{D}_i$ being the Levi--Civita connection associated
with $\tilde{\gamma}_{ij}$. This is in the spirit of the
decomposition employed in the (X)CTS approach to initial data.
Regarding the choice of the representative of the conformal metric
$\tilde{\gamma}_{ij}$, a unimodular condition
$ \tilde{\gamma} = f $ was adopted in~\cite{Bonazzola04},
so that $\psi = \left( \gamma/f \right)^{1/12}$.
The deviation of the conformal metric from the flat fiducial metric
is denoted by  $h^{ij}$, i.e.\
\begin{equation}
  h^{ij} := \tilde{\gamma}^{ij} - f^{ij}.
  \label{e:hij}
\end{equation}
Once the $ 3 + 1 $ conformal decomposition is performed,
a choice of gauge is needed in order to properly reformulate the
Einstein equations as partial differential equations. The
prescriptions in~\cite{Bonazzola04} are \emph{maximal
slicing} and the so-called \emph{generalized Dirac gauge}, 
\begin{equation}
  K = 0,
  \qquad \DSc_k \tilde{\gamma}^{ki} = 0
  \label{e:gauges} ,
\end{equation}
where $\DSc_k$ stands for the 
Levi--Civita connection associated with the flat metric $f_{ij}$. 
The Einstein equations then become a coupled
elliptic-hyperbolic system to be solved for the basic variables
$h^{ij}$, $\psi$, $N$, and $\beta^i$~\cite{Bonazzola04}.

Expressing the differential operators in terms of the connection of
the flat metric, the elliptic part can be written as 
\begin{widetext}
  \begin{equationarray}
    \Delta\psi & = &
    - 2 \pi E \psi^5
     - h^{kl} {\cal D}_k {\cal D}_l \psi +
    \psi \frac{\tilde{R}}{8}
    \nonumber
    \\
    & & 
    - \frac{\psi^5}{8 (2 N)^2}
     \tilde{\gamma}_{ik} \tilde{\gamma}_{jl}
     \!\left[ (L \beta)^{ij}
    \! + \! \frac{\partial h^{ij}}{\partial t} \! - \!
    \Lie{\beta} h^{ij} \! - \! \frac{2}{3} {\cal D}_k \beta^k h^{ij} \right] 
    \!\!
    \left[ (L \beta)^{kl}
     \! + \! \frac{\partial h^{kl}}{\partial t} \! - \!
    \Lie{\beta} h^{kl} \! - \! \frac{2}{3} {\cal D}_m \beta^m h^{kl} \right] \!,
    \label{e:conformal_factor}
    \\ [1 em]
    \Delta (N \psi) & = &
    2 N \psi^5 \pi (E + 2 S)
     + N \psi \frac{\tilde{R}}{8} -
    h^{kl} {\cal D}_k {\cal D}_l (N \psi)
    \nonumber
    \\
    & &
    + \frac{7}{32} \frac{\psi^6}{(N \psi)}
     \tilde{\gamma}_{ik} \tilde{\gamma}_{jl}
    \! \left[ (L \beta)^{ij}
     \! + \! \frac{\partial h^{ij}}{\partial t} \! - \!
    \Lie{\beta} h^{ij} \! - \! \frac{2}{3} {\cal D}_k \beta^k h^{ij} \right]
    \!\!
    \left[ (L \beta)^{kl}
     \! + \! \frac{\partial h^{kl}}{\partial t} \! - \!
    \Lie{\beta} h^{kl} \! - \! \frac{2}{3} {\cal D}_k \beta^k h^{kl} \right] \!, 
    ~~~~
    \label{e:conformal_factor*lapse}
    \\ [1 em]
    \Delta \beta^i + \frac{1}{3} {\cal D}^i {\cal D}_j \beta^j & = &
    16 \pi N \psi^4 S^i
     - h^{kl} {\cal D}_k {\cal D}_l \beta^i -
    \frac{1}{3} h^{ik} {\cal D}_k {\cal D}_l \beta^l +
    \frac{\psi^6}{N} {\cal D}_j \left( \frac{N}{\psi^6} \right)
    \left[ (L \beta)^{ij} \right]
    \nonumber
    \\
    & &
     + \frac{\psi^6}{N} {\cal D}_j \left( \frac{N}{\psi^6} \right)
    \left[ \frac{\partial h^{ij}}{\partial t} - \Lie{\beta} h^{ij} \! -
    \frac{2}{3} {\cal D}_k \beta^k h^{ij} \right] -
    2 N \Delta^i_{kl} \tilde{A}^{kl},
    \label{e:shift}
  \end{equationarray}%
\end{widetext}
where $\Delta$ stands for the flat Laplacian ($\Delta:=f^{ij}{\cal D}_i{\cal D}_j$),
$E$, $S^i$ and $S$ are, respectively, the energy density, momentum density, and trace of the 
stress tensor, all measured by the observer of 4-velocity $n^\mu$ (Eulerian observer):
in terms of the energy-momentum tensor $T_{\mu\nu}$,
$E:=T_{\mu\nu}n^{\mu}n^{\nu}$,
$S^i:=-\gamma^{i\mu}T_{\mu\nu}n^{\nu}$, and 
$S:=\gamma^{ij} S_{ij}$, with 
$S_{ij} := T_{\mu\nu} \gamma^\mu{}_i \gamma^\nu{}_j$. Furthermore,
\begin{equationarray}
  \tilde{R} & = &
  \frac{1}{4} \tilde{\gamma}^{kl} {\cal D}_k h^{mn} {\cal D}_l
  \tilde{\gamma}_{mn} - \frac{1}{2} \tilde{\gamma}^{kl} {\cal D}_k
  h^{mn} {\cal D}_n \tilde{\gamma}_{ml}, \quad \label{e:tildeR}
  \\
  (L \beta)^{ij} & := &
  {\cal D}^i \beta^j + {\cal D}^j \beta^i -
  \frac{2}{3} f^{ij} {\cal D}_k \beta^k, \label{e:def_confKilling}
  \\
  \Delta^k_{ij} & := & \frac{1}{2} \tilde{\gamma}^{kl}
  \left( {\cal D}_i \tilde{\gamma}_{lj} + 
  {\cal D}_j \tilde{\gamma}_{il} -
  {\cal D}_l \tilde{\gamma}_{ij} \right).
\end{equationarray}%
Equation~(\ref{e:conformal_factor}) follows from the Hamiltonian constraint,
whereas Eq.~(\ref{e:shift})  results from the momentum constraint
together with the preservation of the Dirac gauge in time.
Equation~(\ref{e:conformal_factor*lapse}) corresponds to the preservation in time of the
maximal slicing condition, $\partial K / \partial t=0$. 
Note that expression (\ref{e:tildeR}) for the Ricci scalar of the conformal metric does not involve any second order derivative of the metric; this property follows from Dirac gauge 
\cite{Bonazzola04}.
The resulting elliptic subsystem coincides 
with the XCTS system~\cite{PfeYor03}, except from the field chosen to solve 
the maximal slicing equation: Eq.~(\ref{e:conformal_factor*lapse})
above is to be solved for $N\psi$, whereas in~\cite{PfeYor03} the
\emph{conformal lapse} $\tilde{N}:=N\psi^{-6}$ is employed instead. This directly affects the value (and, in particular, the sign) of the power of
the conformal factor in the nonlinear terms of 
Eqs.~(\ref{e:conformal_factor}) and (\ref{e:conformal_factor*lapse}). 
More generally, one could define a
generic rescaling of the lapse, $N=\tilde{N}\psi^a$, such that the choice in~\cite{PfeYor03}
corresponds to $a=6$, whereas the choice in Eq.~(\ref{e:conformal_factor*lapse}) above 
corresponds to $a=-1$ (see~\cite{JarAnsLim07} for the general equations
in the vacuum case). An important remark is the absence of a 
choice of $a$ such that the factors multiplying
$\psi$ and $\tilde{N}$ on the right hand side of the 
linearized versions of Eqs.~(\ref{e:conformal_factor})
and~(\ref{e:conformal_factor*lapse}) both present a
positive sign. 
In the presence of matter, terms multiplying the energy density $E$ 
also contribute to these sign difficulties, though in this case they can be fixed
by an appropriate conformal rescaling of the energy density (see later).
An additional concern in a generic evolution
scenario is the sign of $\tilde{R}$, which is also relevant
in the linearized equations. Implications of this
issue are discussed in Sec.~\ref{sect:schemeCFC}.

The Einstein equations in the form of the elliptic
equations~(\ref{e:conformal_factor})-(\ref{e:shift}) 
and the hyperbolic equation for $h^{ij}$ as given in Ref.~\cite{Bonazzola04}  
are to be solved together with the hydrodynamic equations,
\begin{equationarray}
  \nabla_{\mu} (\rho u^{\mu}) & = & 0,
  \label{e:hydro1}
  \\
  \nabla_{\mu} T^\mu{}_\nu & = & 0,
  \label{e:hydro2},
\end{equationarray}%
where $\nabla^{\mu}$ is the Levi--Civita connection
associated with the metric $g_{\mu\nu}$, $\rho$ is the rest-mass (baryon mass)
density, and $u^{\mu}$ is the 4-velocity of the fluid.


\subsection{The conformal flatness approximation}
\label{subsection:cfc_approximation}

If the hyperbolic part of the FCF system is not solved, but rather the 
condition $h^{ij}=0$ is imposed, the
resulting three-metric $\gamma_{ij}$ is conformally flat, and the CFC approximation is
recovered. Therefore, the FCF is a natural generalization of the
CFC approximation. The latter has been used in many
astrophysical applications, like the rotational collapse of cores
of massive stars~\cite{Dimmelmeier02b, Dimmelmeier05, ott_07_a,
  cerda_07_a} or supermassive stars~\cite{saijo_04_a}, the
phase-transition-induced collapse of rotating neutron stars to
hybrid quark stars~\cite{abdikamalov_08_a}, equilibrium models of
rotating neutron stars~\cite{cook_96_a, Dimmelmeier06}, as well as for
binary neutron star merger~\cite{Wilson96, oechslin_02_a, Oechslin07,
  faber_04_a}. The elliptic subsystem 
of the FCF,
Eqs.~(\ref{e:conformal_factor})--(\ref{e:shift}), reduces in CFC to
\begin{equationarray}
  \Delta\psi & = &
  - 2 \pi \psi^{-1}
  \! \left[ E^* \! + \! \frac{\psi^6 K_{ij} K^{ij}}{16 \pi} \right] \!\!,
  \label{e:psi_cfc0}
  \\
  \Delta (N \psi) & = &
  2 \pi N \psi^{-1}
  \! \left[E^* \! + \! 2 S^* \! + \! \frac{7 \psi^6 K^{ij} K_{ij}}{16 \pi} \right] \!\!,
  ~~~~~\,
  \label{e:Npsi_cfc0}
  \\
  \Delta \beta^i \!\! + \! \frac{1}{3}{\cal D}^i {\cal D}_j \beta^j & = &
  16 \pi N \psi^{-2} (S^*)^i \! + 2 \psi^{10} K^{ij} {\cal D}_j
   \frac{N}{\psi^6} ,
  \label{e:beta_cfc0}
\end{equationarray}%
where the following rescaled matter quantities have been introduced,
following York~\cite{York79}:
\begin{equationarray}
  E^* & := \sqrt{\gamma/f} \, E & = \psi^6 E,
  \label{e:def_Estar}
  \\
  S^* & := \sqrt{\gamma/f} \, S & = \psi^6 S,
  \\
  (S^*)_i & := \sqrt{\gamma/f} \, S_i & = \psi^6 S_i.
  \label{e:def_Sistar}
\end{equationarray}%
Equations~(\ref{e:psi_cfc0})
and~(\ref{e:Npsi_cfc0}) inherit the
local nonuniqueness problems already present in the
FCF equations. Although the sign problems specifically related to the energy density terms 
are solved by the conformal rescaling of the components of the energy-momentum tensor 
and the CFC eliminates the $\tilde{R}$ term,
problems related to the $K_{ij}K^{ij}$ term remain in the scalar CFC
equations. This is apparent once the extrinsic curvature is expressed
in terms of the lapse and the shift.

Conformal rescaling of the hydrodynamical variables is not only relevant
for local uniqueness issues. The hydrodynamic equations~(\ref{e:hydro1}) and 
(\ref{e:hydro2}) can be
formulated as a first-order hyperbolic system of
conservation equations for the quantities
$(D^*, (S^*)_i, E^*)$~\cite{Banyuls97, Font07}, where, similarly to 
Eqs.~(\ref{e:def_Estar})--(\ref{e:def_Sistar}), 
$D^*:= \psi^6 D$, $D:=N u^0 \, \rho$ being the baryon mass density
as measured by the Eulerian observer. We can thus consider $E^*$ and $(S^*)_i$ as  
known variables in the computation of the CFC
metric. Note that these quantities differ from $E$ and $S_i$ by a
factor $\psi^6$, and hence it is not possible to compute the nonstarred
quantities before knowing the value of $\psi$.
If the energy-momentum tensor represents a fluid, then the source of
Eq.~(\ref{e:Npsi_cfc0}) cannot be explicitly expressed in terms of
$(D^*, (S^*)_i, E^*)$, the reason for that
being the dependence of $S^*$ on the pressure $P$. The pressure can only
be computed in terms of the ``primitive'' quantities, e.g., as a
function $P(\rho,\epsilon)$ of the rest-mass density and the specific
internal energy $\epsilon$. The primitive quantities are, in general,
recovered from $(D, S_i, E)$ implicitly by means of an iteration
algorithm. So far, two solutions of the problem related to the fact
that $S^*$ directly contains $ P $ have been used in numerical simulations
performed with the CFC approximation.

The first approach~\cite{Wilson96} is to consider $P$, and hence also
$S^*$, as an implicit function of $\psi$. Then
Eqs.~(\ref{e:psi_cfc0})--(\ref{e:beta_cfc0}) can be solved as a
coupled set of nonlinear equations using a fixed-point iteration
algorithm. The convergence of the algorithm to the correct solution
depends not only on the proximity of the initial seed metric to the solution,
but also on the uniqueness of this solution. The latter point is
extensively discussed in Sec.~\ref{sect:schemeCFC}. Furthermore, one problem of
this approach is the necessity of performing the recovery of the
primitive variables (which is numerically a time consuming procedure)
to compute the pressure during each fixed point iteration. Because of the
uniqueness problem, this approach
can be only successfully applied in numerical simulations for, at most,
moderately strong gravity (like stellar core collapse to a neutron
star or the inspiral and initial merger phase of binary neutron stars), but fails for
more compact configurations like the collapse of a stellar core or a
neutron star to a black hole. For such scenarios with very strong
gravity, one finds convergence of the metric to a physically
incorrect solution of the equations or even nonconvergence of the
algorithm.

A second approach to the recovery algorithm problem is the attempt to
calculate $P$ independently of the CFC equations. This can be achieved
by computing the conformal factor by means of the evolution equation
\begin{equation}
  \der{\psi'}{t} = \frac{\psi'}{6} {\cal D}_k \beta^k.
  \label{e:der_conf_fact}
\end{equation}
The conformal factor $\psi'$ obtained in this way is analytically
identical to the $\psi$ from
Eqs.~(\ref{e:psi_cfc0})--(\ref{e:beta_cfc0}), but here we use a
different notation to keep track of the way it is computed. The value
of $\psi'$ is solely used to evaluate $P$, and the coupled system of
Eqs.~(\ref{e:psi_cfc0})--(\ref{e:beta_cfc0}) is solved for
determining $\psi$, $N$, and $\beta^i$. Although this approach allows
one to avoid the problem of recovering the primitive variables at each
iteration, it also suffers from the convergence problem, and the
simulation of configurations with very strong gravity is still not
feasible. Furthermore, new complications
are introduced by using two differently computed values,
$\psi$ and $\psi'$, of the same quantity. For some scenarios like the
formation of a black hole from stellar collapse, the numerical values
of these two quantities during the evolution of the system start to
diverge significantly at some point. We find that this
inconsistency cannot be avoided, since any attempt to artificially
synchronize both values leads to numerical instabilities.


\section{The new scheme in the conformally flat case}
\label{sect:schemeCFC}


\subsection{Uniqueness of the elliptic equations and convergence of
  elliptic solvers}

Well-posed elliptic partial differential systems admit non-unique solutions
whenever the associated differential operator has
a nontrivial kernel. When discussing sufficient conditions guaranteeing 
uniqueness, it is illustrative to first consider the case of a scalar elliptic equation.
In particular, for the class of scalar elliptic equations for the function $u$ of the form
\begin{equation}
  \Delta u + h u^p = g,
  \label{e:sign}
\end{equation}
where $h$ and $g$ are known functions independent of $u$, a 
maximum principle can be used to prove
local uniqueness of the solutions as long as the sign of the
exponent $p$ is different from the sign of the proper function 
$h$~\cite{Protter67, York79, Taylo96, Evans98}.

In the CFC case, we are not dealing with a single scalar elliptic
equation, but rather with the coupled non-linear elliptic
system~(\ref{e:psi_cfc0})-(\ref{e:beta_cfc0}).
Therefore, assessing wheter or not the scalar equations~(\ref{e:psi_cfc0})
and~(\ref{e:Npsi_cfc0}) present the good signs for the application
of a maximum
principle is an important step for understanding
the uniqueness properties of the whole system.
However, as pointed out in the previous section, the
CFC equations for the conformal factor and the lapse
possess the wrong signs in the quadratic
extrinsic curvature terms (once everything is expressed in terms
of the lapse and the shift). This problem
can be fixed in Eq.~(\ref{e:psi_cfc0}) 
by an appropriate rescaling of the lapse, $N=\tilde{N}\psi^6$,
but this strategy does not solve the problem for the lapse equation
(cf.\ the discussion on the conformal lapse $\tilde{N}$ in Sec.~\ref{ss:brief_review}).
Therefore, we cannot use the maximum principle to infer
local uniqueness of the solutions to the CFC equations. 
In these conditions of potential nonunique solutions, convergence to a 
undesirable solution may happen. As mentioned in the introduction,
this pathology has been illustrated using simple analytical
examples of scalar equations of the type~(\ref{e:sign})
in~\cite{BauOMuPfe07}, as well as in numerical implementations of the vacuum Einstein
constraints in the XCTS approach~\cite{Pfeiffer05} and certain
constrained evolution formalisms (see, e.g., \cite{Rinne08}).

In the context of the CFC approximation this sign issue has also appeared,
in particular associated with the ``recovery algorithm'' problem discussed
in Sec.~\ref{subsection:cfc_approximation} since it involves the evaluation
of the conformal factor. 
Nonunique solutions of $\psi$, either due to the use of the
nonconformally rescaled $E$ or the quadratic extrinsic curvature term,
spoil the convergence of the algorithm when density, and 
thus compactness, increases. We again emphasize that a possible
synchronization of $\psi$ and $\psi'$ does not solve the problem in
general, since numerical instabilities eventually arise at
sufficiently high compactness.


\subsection{Numerical examples}

\begin{figure}[t]
  \epsfxsize = 8.6 cm
  \centerline{\epsfbox{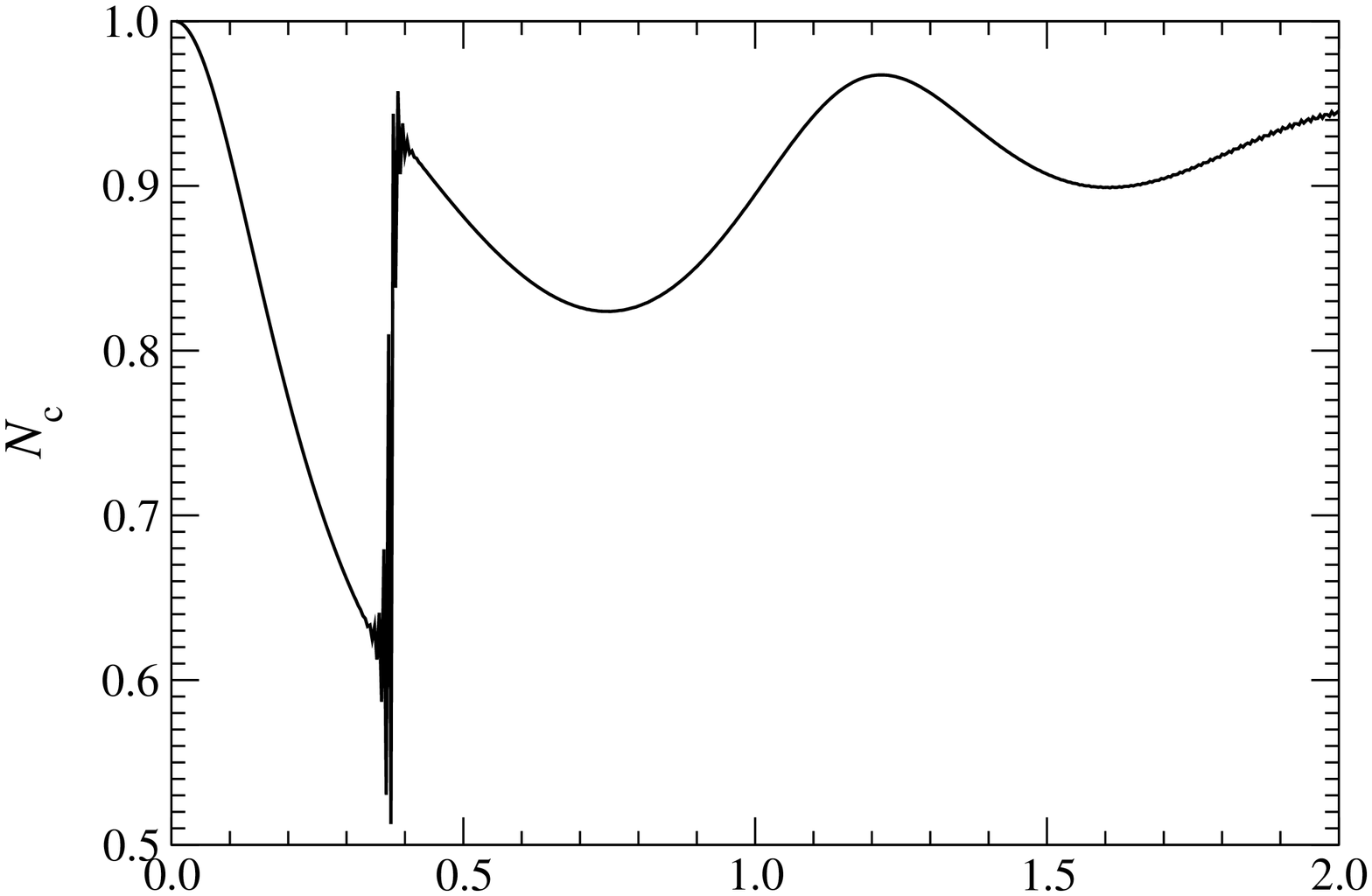}}
  \vspace{1 em}
  \epsfxsize = 8.6 cm
  \centerline{\epsfbox{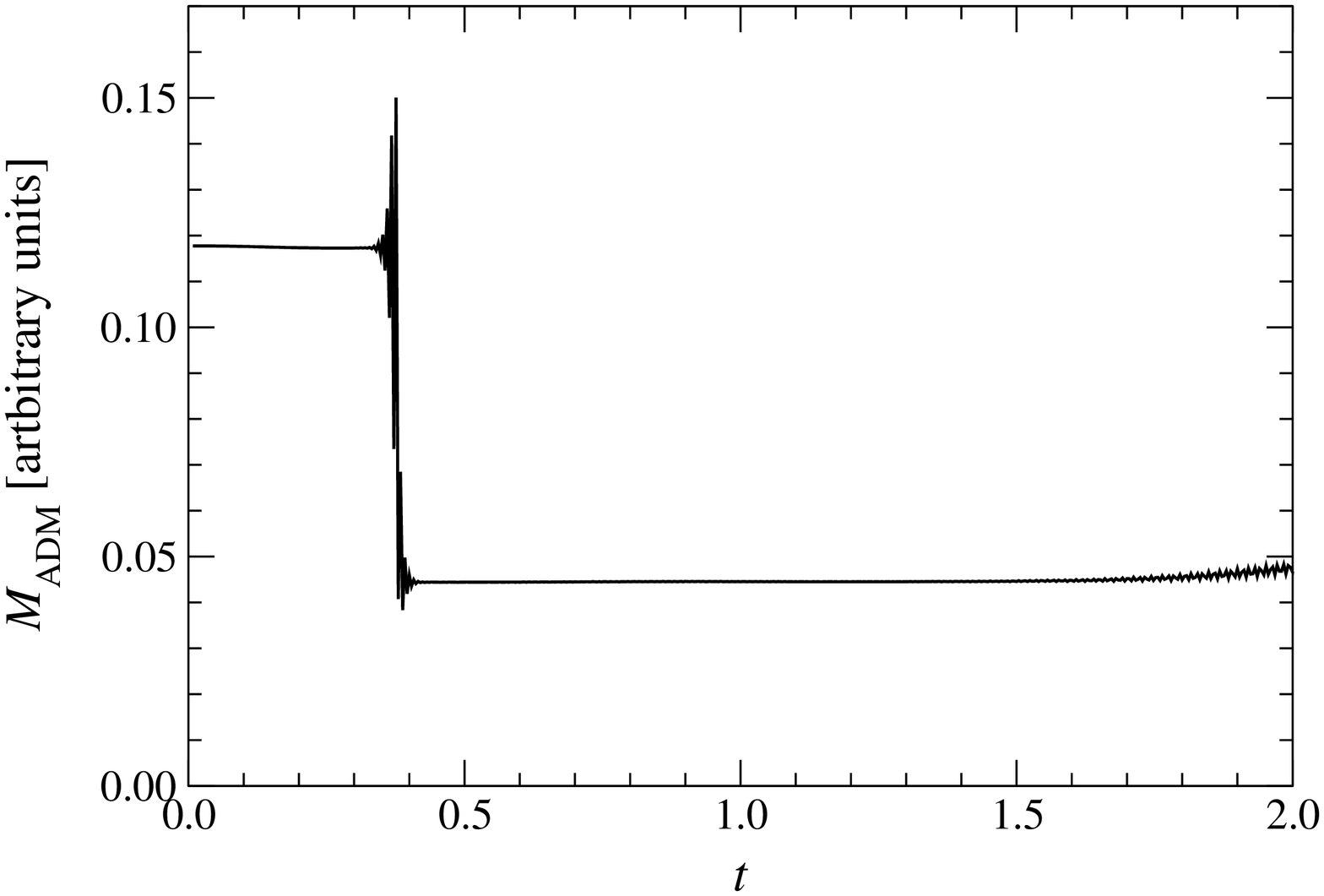}}
  \caption{Time evolution of the central lapse $ N_\mathrm{c} $ (top
    panel) and the ADM mass $ M_\mathrm{ADM} $ (bottom panel) for
    a collapsing packet of gravitational waves, using the 
    integration scheme proposed in~\cite{Bonazzola04}. The unit of $t$ is given by the initial width of the wave packet}
  \label{f:FCF_wave}
\end{figure}

The nonuniqueness of solutions has also been observed in FCF, 
as described in the following example. Let us consider a
vacuum spacetime, with initial data formed by a Gaussian wave packet, as
in~\cite{Bonazzola04}, but with much higher amplitude $\chi_0 = 0.9$ instead of 
$\chi_0=10^{-3}$ in~\cite{Bonazzola04} (see the latter reference for notations). 
The integration technique and
numerical settings are the same as in~\cite{Bonazzola04}, but contrary to
the results for small amplitudes obtained in that reference, the wave packet does not 
disperse to infinity and instead starts to collapse. 
Fig.~\ref{f:FCF_wave} displays the time evolution of the 
central lapse $ N_\mathrm{c} $ at $ r = 0 $ and of the system's 
Arnowitt-Deser-Misner (ADM) mass
$ M_\mathrm{ADM} $, which in the present conformal
decomposition can be expressed as 
\begin{equationarray}
  M_\mathrm{ADM} & = &
  - \frac{1}{2 \pi} \oint_\infty \left( {\cal D}_i \psi - \frac{1}{8}
  {\cal D}^j \tilde{\gamma}_{ij} \!\right) d{\cal A}^i
  \nonumber \\
  & = & - \frac{1}{2 \pi} \oint_\infty {\cal D}_i \psi \, d{\cal A}^i,
  \label{e:ADM_mass}
\end{equationarray}%
where the integral is taken over a sphere of radius $ r = \infty $
and the second equality follows from the use of Dirac gauge [Eq.~(\ref{e:gauges})].

The very sudden change at $t\simeq 0.4$ in both the central lapse and the ADM
mass, which is also present in, e.g., the central conformal factor
$\psi_\mathrm{c}$, originates from the convergence of the elliptic
system~(\ref{e:conformal_factor})-(\ref{e:shift}) to another solution with a
different (unphysical) value of the ADM mass. The good conservation of $
M_\mathrm{ADM} $ and the smooth evolution of $N_\mathrm{c}$ for $t\gtrsim 0.4$
indicate that this other solution remains stable until $t\simeq 2$, when
high-frequency oscillations appear. These oscillations may be due to the
overall inconsistency of the system, destabilizing the whole scheme. On the
other hand, the time evolution of $h^{ij}$ does not show any such type of
behavior, and $h^{ij}$ exhibits a continuous radial profile at all times. This
is numerical evidence that, also for the full Einstein case (i.e.\ without
approximation), the generalized elliptic equations suffer from a similar
convergence problem as in the CFC case.

The same subject is also exemplified when one tries to calculate the
spacetime metric for an equilibrium neutron star model from the
unstable branch using either
Eqs.~(\ref{e:conformal_factor})--(\ref{e:shift}) in the FCF case or
Eqs.~(\ref{e:psi_cfc0})--(\ref{e:beta_cfc0}) in the CFC
approximation. Even for the simple setup of a polytrope with adiabatic
index $ \Gamma = 2 $ in spherical symmetry, those metric equations
yield -- when converging at all -- a grossly incorrect solution if the
matter quantities $(D, S_i, E)$ in the source terms are held fixed.
Both the metric components as well as the ADM mass can deviate from
the physical solution by a few tens of percent, even though that
incorrect metric satisfies the asymptotic flatness condition. The
reason why programs for constructing rotating relativistic neutron
star models, like the KEH code~\cite{komatsu_89_a}, the RNS
code~\cite{stergioulas_95_a}, or the BGSM code~\cite{bonazzola_93_a},
are not obstructed by this nonuniqueness problem is apparently that
they all utilize an iteration over both the metric and the hydrodynamic
equations \emph{simultaneously}, thereby allowing the matter
quantities to change during the calculation of the metric.

We want to stress here that these non-convergence issues in the CFC
case are not related to the approximation that is made. If one
considers this system in the spherical (one-dimensional) case,
CFC is no longer an approximation, but is the choice of the so-called
isotropic gauge. Even then, the elliptic
system~(\ref{e:psi_cfc0})-(\ref{e:beta_cfc0}) no longer converges
to the proper (physical) solution.


\subsection{The new scheme and its theoretical properties}
\label{s:new_scheme}

Despite the above mentioned convergence problems, numerically simulating 
the physical problem of
spherical collapse to a black hole in isotropic coordinates has been
successfully studied by Shapiro and Teukolsky in~\cite{Shapiro85}. Because 
of the spherical symmetry, 
there exists only one independent component of the extrinsic curvature. It
is then possible to compute directly a conformal extrinsic curvature, $\psi^6
K^r_r$, from the conserved hydrodynamical variables. The elliptic equation for
$\psi$ then decouples from the other elliptic equations by
introducing this conformal extrinsic curvature and using the conserved
hydrodynamical variables in the source. This source term presents no
problem for proving local uniqueness, and
the equation for $\psi$ always converges to the physically correct solution. Once the conformal
factor, the extrinsic curvature (from the conformal factor and the conformal
extrinsic curvature), and the conserved hydrodynamical variables are known, the
elliptic equation for $N\psi$ can be solved and, again, the source
exhibits no local uniqueness problem. This follows
from the fact that the extrinsic curvature is not expressed
in terms of the lapse and the shift. This contrasts with the
CFC equation~(\ref{e:Npsi_cfc0}) where a division
by $N^2$ occurs in the last term when the extrinsic curvature is
expressed in terms of its constituents $ N $, $ \psi $, and $ \beta^i $.
In addition, there is no need to use $\psi'$. Finally, the elliptic equation for the
shift vector can be solved. In summary, no problems of instabilities
or divergence are encountered.

We now generalize this scheme to the CFC case in three
dimensions. This involves the use of two different conformal decompositions 
of the extrinsic curvature: first, two different
conformal rescaling and, second, two different decompositions
of the traceless part into longitudinal and transverse parts.
Adopting maximal slicing, $K=0$, a generic conformal decomposition can
be written as
\begin{equation}
  K^{ij} = \psi^{\zeta-8} (A^{(\zeta)})^{ij} :=
  \psi^{\zeta-8} \left( \frac{1}{\sigma} (L X)^{ij} + A_{\mathrm{TT}}^{ij} \right),
  \label{e:conf_decomp}
\end{equation}
where $\zeta$ is a free parameter and $\sigma$ a free function,
$A_{\mathrm{TT}}^{ij}$ is transverse traceless and $L$ in the conformal
Killing operator defined by Eq.~(\ref{e:def_confKilling}). We implicitly make
use of a flat conformal metric, with respect to which $A_{\mathrm{TT}}^{ij}$ is
transverse, although, in principle, it would be more general to use the metric
$\tilde{\gamma}^{ij}$ and the conformal Killing operator associated with it,
$\tilde{L}$. But such a decomposition would introduce many technical
difficulties in our treatment.
In particular, it is numerically easier to handle tensors which are
divergence-free with respect to the flat metric in the generalization to FCF.
The vector $X^i$, on which $L$ is acting, is therefore called the \emph{longitudinal
  part} of $(A^{(\zeta)})^{ij}$.  The first decomposition we use is the one
introduced in Eqs.~(\ref{e:conformal_decomp}) and (\ref{e:Aij}) with the choice
$\zeta=4$ and $\sigma=2N$. This corresponds to a CTS-like decomposition of the
traceless part, so that $X^i$ is given by the shift vector $\beta^i$ and
$A_{\mathrm{TT}}^{ij}$ can be expressed in terms of the time derivative of the conformal metric. We
denote this traceless part as $\tilde{A}^{ij}:= (A^{(4)})^{ij}$. In the CFC
approximation this becomes
\begin{equation}
  K^{ij} = \psi^{-4} \tilde{A}^{ij},
  \qquad
  \tilde{A}^{ij} = \frac{1}{2 N} (L \beta)^{ij}.
  \label{e:tildeAij}
\end{equation}
The second conformal decomposition,
\begin{equation}
  K^{ij} = \psi^{-10} \hat{A}^{ij},
  \qquad
  \hat{A}^{ij} = (L X)^{ij} + {\hat A}_{\mathrm{TT}}^{ij},
  \label{e:hatdefAij}
\end{equation}
refers to $\zeta=-2$  and $\sigma=1$. It instead
corresponds to a conformal transverse traceless (CTT) decomposition of
the traceless part of extrinsic curvature introduced by Lichnerowicz~\cite{Lichn44}.
Notice that we have defined $\hat{A}^{ij}:= (A^{(-2)})^{ij}$, not to be confused 
with $\tilde{A}^{ij}:= (A^{(4)})^{ij}$. The relation between
$\hat{A}^{ij}$ and $\tilde{A}^{ij}$ is given by
\begin{equation}
  \hat{A}^{ij} = \psi^{10} K^{ij} = \psi^6 \tilde{A}^{ij}.
  \label{e:hatAij}
\end{equation}
In terms of $\hat{A}^{ij}$, the CFC momentum constraint can be written as
\begin{equation}
  {\cal D}_j \hat{A}^{ij} =
  8 \pi \psi^{10} S^i = 8 \pi \psi^6 f^{ij} S_j = 8 \pi f^{ij} S^*_j.
  \label{e:div_hat_A}
\end{equation}
Consistency between the CTT-like decomposition~(\ref{e:hatdefAij}) and the
CTS-like one~(\ref{e:tildeAij}) generically requires a
nonvanishing tranverse part ${\hat A}_{\mathrm{TT}}^{ij}$ in
Eq.~(\ref{e:hatdefAij}). However, as it is shown in the
Appendix, this ${\hat A}_{\mathrm{TT}}^{ij}$ is smaller in
amplitude than the nonconformal part $h^{ij}$ of the spatial metric and
$\hat{A}^{ij}$ can be approximated on the CFC approximation level as
\begin{equation}
  \hat{A}^{ij} \approx (LX)^{ij} =
  {\cal D}^i X^j + {\cal D}^j X^i - \frac{2}{3} {\cal D}_k X^k f^{ij}.
  \label{e:decomp_tA}
\end{equation}
From Eqs.~(\ref{e:hatdefAij}) and (\ref{e:div_hat_A}),  an elliptic equation for
the vector $X^i$ can be derived,
\begin{equation}
  \Delta X^i + \frac{1}{3} {\cal D}^i {\cal D}_j X^j = 8 \pi f^{ij} S^*_j,
  \label{e:equ_X}
\end{equation}
from which $X^i$ can be obtained. With this vector field, one can calculate
the tensor $\hat{A}^{ij}$ via (\ref{e:decomp_tA}). Notice that in the case of spherical symmetry,
$\hat{A}^{rr}=\psi^{10}K^{rr}=\psi^6K^r_{\ r}$ is the quantity used by
Shapiro and Teukolsky~\cite{Shapiro85}.

The elliptic equation for the conformal factor can be rewritten in terms of
the conserved hydrodynamical variables and $\hat{A}^{ij}$:
\begin{equationarray}
  \Delta \psi = - 2 \pi \psi^{-1} E^* -
  \psi^{-7} \frac{f_{il} f_{jm} \hat{A}^{lm} \hat{A}^{ij}}{8}.
  \label{e:psiCFC}
\end{equationarray}%
This equation can be solved in order to obtain the conformal factor.
Once the conformal factor is known, the procedure to implicitly
recover the primitive variables from the conserved ones is possible,
the pressure $P$ can be computed using the equation of state, and therefore
$S^*$ is at hand. The elliptic equation for $N\psi$ can be
reformulated by means of the conserved hydrodynamical variables,
$\hat{A}^{ij}$, and the conformal factor:
\begin{equationarray}
  \Delta (\psi N) = 2 \pi N \psi^{-1} \left( E^* + 2 S^* \right) +
  N \psi^{-7} \frac{7 f_{il} f_{jm} \hat{A}^{lm} \hat{A}^{ij}}{8}.
  \nonumber \\
  \label{e:NpsiCFC}
\end{equationarray}%
From this equation $N\psi$ can then be obtained and, consequently, so can the
lapse function $N$. Note that, since $\hat{A}^{ij}$ is already known at this step, no
division by $N^2$ spoils the good sign for the maximum principle.

Using the relation between the two conformal decompositions of the
extrinsic curvature, $\hat{A}^{ij}=\psi^6\tilde{A}^{ij}$, Eq.~(\ref{e:tildeAij})
can be expressed as
$\left(L\beta\right)^{ij}=2N\psi^{-6}\hat{A}^{ij}$. Taking the 
divergence, we arrive at an elliptic equation for the shift
vector,
\begin{equation}
  \Delta \beta^i + \frac{1}{3} {\cal D}^i \left({\cal D}_j \beta^j \right) =
  {\cal D}_j \left( 2 N \psi^{-6} \hat{A}^{ij} \right),
  \label{e:shiftCFC}
\end{equation}
where the source is completely known. This elliptic equation can be solved in
order to obtain the shift vector $\beta^i$ consistent with 
$\partial_t\tilde{\gamma}_{ij}=0$, as required by the CFC approximation.

In this recast of the CFC equations, an extra elliptic vectorial
equation for the vector field $X^i$ is introduced. However, now the 
signs of the exponents of $ \psi $ and $ N $ are compatible with the
maximum principle for scalar elliptic equations, and
the problem is \emph{linearization stable}. While this does not guarantee
\emph{global} uniqueness of the solutions, it provides a sufficient
result for \emph{local} uniqueness. 
This strongly relies on the fact that the system decouples in a
hierarchical way, which we summarize here once more:
\begin{enumerate}
\item With the hydrodynamical conserved quantities at hand, solve
  Eq.~(\ref{e:equ_X}) for $X^i$, and thus for $\hat{A}^{ij}$.
\item Solve Eq.~(\ref{e:psiCFC}) for $\psi$, where local uniqueness
  is now guaranteed. Then $ S^* $ can be calculated consistently.
\item Solve Eq.~(\ref{e:NpsiCFC}) for $N\psi$, a linear equation
  where the maximum principle can be applied and uniqueness and
  existence follow with appropriate boundary conditions. 
\item As the source of Eq.~(\ref{e:shiftCFC}) is then fully known,
  solve it for $\beta^i$.
\end{enumerate}
Note that this scheme is similar to that used by Shibata and
Ury\=u~\cite{ShibaU06} to compute initial data for black hole -
neutron star binaries. We will discuss this point further in
Sec.~\ref{s:compar_prev_works}.

The new CFC metric equations presented here not only allow us to evolve
the hydrodynamical equations and recover the metric variables
from the elliptic equations in a consistent way (no auxiliary quantity $\psi'$ is needed), 
but they also permit to
introduce initial perturbations in the hydrodynamical variables
(strictly speaking, in the conserved quantities) in a set of previously
calculated initial data and directly delivers the correct values for
the metric. It is even possible to perturb only the primitive
quantities, and consistently resolve for the metric by iterating until
the conformal factor $ \psi $, which links the primitive to the
conserved quantities, converges. We find that such an iteration method
fails for sufficiently strong gravity if the original CFC formulation
is used.


\section{Numerical results}
\label{s:numerical_results}

We recapitulate that the original CFC formulation exhibits serious
convergence problems when dealing with highly compact configurations
such as nascent black holes.
This weakness of the original formalism is noticeable in the fact that no
simulations of rotational collapse to a black hole substantially
beyond the formation of the apparent horizon have been
performed so far in the CFC. Furthermore, some scenarios which do
not involve the formation of a black hole are alredy feasible with the old
formulation only if procedures like using Eq.~(\ref{e:der_conf_fact}),
with all associated problems and inconsistencies, are employed. An
example is the migration of a neutron star model from the stable to
the unstable branch, which is a standard test for relativistic
hydrodynamics codes. In contrast, the new CFC scheme presented in
this work solves all problems that prevented performing such
simulations in the past. In order to show the suitability of the new
scheme we present the results of numerical simulations of the
migration test and of the rotational collapse to a black hole.


\subsection{Model setup}

The numerical simulations presented here are performed using the
numerical code CoCoNuT~\cite{Dimmelmeier02a, Dimmelmeier05}. This code
solves the evolution of the hydrodynamics equations coupled to the
elliptic equations for the spacetime metric in the CFC
approximation. Standard high-resolution
shock-capturing schemes are used in the hydrodynamic evolution, while
spectral methods are employed to solve the metric equations. The code
is based on spherical polar coordinates, and for the tests presented here we
assume axisymmetry and symmetry with respect to the equatorial plane.
Note that the metric equations presented in this paper are covariant. Thus
the formalism can be used for any coordinate basis as well as
without any symmetry conditions.

The initial models are general relativistic $\Gamma=2$ polytropes in
equilibrium with a polytropic constant $K=100$. The models are chosen to be
situated on the unstable branch, i.e.\ $\partial M_\mathrm{ADM} / \partial
\rho_\mathrm{c} < 0$, where $ \rho_\mathrm{c} $ is the central rest-mass
density. Therefore, any perturbation of the star induces either the collapse
to a black hole or migration to a configuration of the same baryon mass on the
stable branch. Table~\ref{tab:ini_models} shows the main features of these
initial models. Models D1 to D4 are uniformly rotating models which are
identical to those presented in~\cite{Baiotti05}. The model labeled SU is a
spherical model, while model labeled SS is the counterpart model with the same baryon mass
but it is located on the stable branch. The equilibrium rotating star models in
Dirac gauge (the axisymmetric and stationary limit of FCF) used here are
described in~\cite{Lin06}, and are computed using the
\textsc{Lorene}~\cite{Lorene} library. We map the hydrodynamic and metric quantities to the CFC
code neglecting the $h^{ij}\sim 10^{-3}$ terms, which are negligible due to
their smallness. Alternatively, we compute CFC equilibrium initial models. In
this case we find that the differences with respect to the FCF models are
small ($\sim 0.1 \%$) for representative metric and hydrodynamic
quantities, initially and during the evolution, and therefore we discuss
only the FCF initial models here.

\begin{table}
  \caption{Initial models used in the migration test and the
    rotational collapse to a black hole. $ \rho_\mathrm{c,i} $ is the
    initial central rest-mass density, $ \Omega_\mathrm{i} $ is the
    initial angular velocity, $ r_\mathrm{p,i} / r_\mathrm{e,i}$ is
    the initial ratio of polar to equatorial coordinate radius,
    $ M_\mathrm{ADM} $ is the gravitational ADM mass, and $ J $ is the
    total angular momentum (which is conserved in CFC during the
    evolution in the axisymmetric case). 
    Units in which $G = c = M_{\odot} = 1$ are used.}
  \label{tab:ini_models}
  \begin{ruledtabular}
    \begin{tabular}{lcccccc}
      Model &
      $ \rho_\mathrm{c,i} $ &
      $ \Omega_\mathrm{i} $ &
      $ r_\mathrm{p,i} / r_\mathrm{e,i} $ &
      $ r_\mathrm{e,i} $ &
      $ M_\mathrm{ADM} $ &
      $ J / M_\mathrm{ADM}^2 $ \\
      & [$ 10^{-3} $] & [$ 10^{-2} $] & & &  \\ [0.2 em]
      \hline \rule{0 em}{1.2 em}%
      SU & 8.000 & 0    & 1.00 & 4.267 & 1.447 & 0     \\
      SS & 1.346 & 0    & 1.00 & 7.999 & 1.424 & 0     \\ [0.5 em]
      D1 & 3.280 & 1.73 & 0.95 & 5.947 & 1.665 & 0.207 \\
      D2 & 3.189 & 2.88 & 0.85 & 6.336 & 1.727 & 0.362 \\
      D3 & 3.134 & 3.55 & 0.75 & 6.839 & 1.796 & 0.468 \\
      D4 & 3.116 & 3.95 & 0.65 & 7.611 & 1.859 & 0.542 \\
    \end{tabular}
  \end{ruledtabular}
\end{table}

The hydrodynamic equation are discretized on the finite difference grid with
$n_{r}\times n_{\theta}$ grid points. The radial grid size is $\Delta
r_0$ for the innermost cell and increases geometrically outwards,
while the angular grid is equidistantly spaced. The metric equations
are solved on a spectral grid consisting of $n_\mathrm{d}-1$ radial
domains distributed such as to homogeneously cover the finite difference grid and a
compactified exterior domain extending to radial infinity. On the
spectral grid we resolve each radial domain with $33$ collocation
points. The spherical model needs only one angular collocation point,
while we use $17$ angular points for the rotating models.

We track the location of the apparent horizon by means of a three-dimensional
spectral apparent horizon finder, described in detail and tested
in~\cite{Lin07}. The apparent horizon location is given by a function
$\mathcal{H}(r, \theta)$, which is decomposed into a set of spherical
harmonics. The coefficients of $\mathcal{H}$ in this basis are computed
iteratively, in order to satisfy the condition that the expansion in the 
outgoing null direction vanishes at the apparent horizon location.


\subsection{Migration of unstable neutron stars to the stable branch}

The first test we consider is the migration of a neutron star
model in equilibrium from the unstable branch to the stable branch,
which is a standard but still demanding test for general relativistic
hydrodynamics codes, as it involves the dynamic transition between two
very compact equilibrium states. This test has been performed in the
past in full general relativistic simulations~\cite{Font02}. We start
the evolution with the nonrotating equilibrium model labeled SU. Since
it belongs to the unstable branch, any perturbation from exact
equilibrium (which can, for instance, be caused by discretization errors) 
leads either to a collapse or to an expansion to a new equilibrium 
configuration of the same baryon mass on the stable branch. The corresponding 
equilibrium configuration with the same baryon mass, model SS, has smaller
ADM mass than the initial system (see
Table~\ref{tab:ini_models}). Therefore, to preserve the ADM mass,
the final configuration cannot be exactly the equilibrium model given by SS. The 
energy difference between models SU and SS should be transformed into 
kinetic energy, remaining in the final object in the form of pulsations.

In our case the numerical truncation error is sufficient to trigger the
migration. Since the final neutron star on the stable branch is larger
than the initial model (see Table~\ref{tab:ini_models}), the outer
boundary of the finite difference grid is chosen to be $4.5$ times the
radius of the model SS. We perform two simulations on a finite
difference grid with $150$ or $300$ radial cells and $\Delta r_0 =
0.022$ or $0.012$, respectively. We use $n_\mathrm{d}=6$ radial domains for the
spectral grid. We evolve the system with either a polytropic or
an ideal gas equation of state.

\begin{figure}[t]
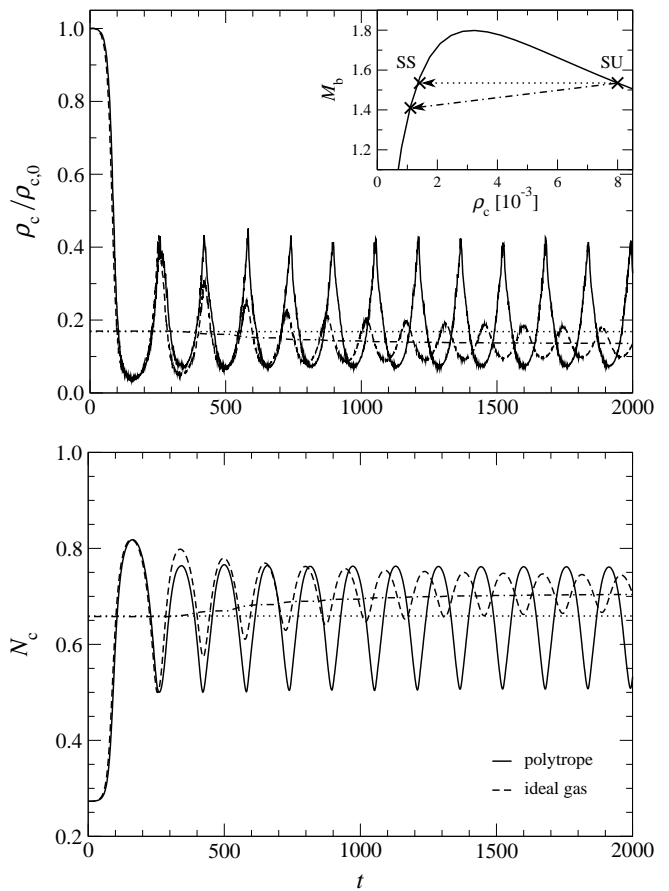

  \epsfxsize = 8.6 cm
  \centerline{\epsfbox{mig_rho.eps}}
  \vspace{1 em}
  \epsfxsize = 8.6 cm
  \centerline{\epsfbox{mig_alpha.eps}}
  \caption{Time evolution of the central rest-mass density
    $ \rho_\mathrm{c} $ (top panel) and the central lapse
    $ N_\mathrm{c} $ (bottom panel) for the migration of the unstable
    neutron star model SU to the stable branch, with either a
    polytropic (solid lines) or an ideal gas (dashed lines) equation
    of state. The dotted horizontal lines mark the value of
    $ \rho_\mathrm{c} $ and $ N_\mathrm{c} $ for the equilibrium
    configuration SS from the stable branch with the same
    baryon mass $ M_\mathrm{b} $ as model SU, while the
    dash-dotted lines are obtained from a series of equilibrium models
    where mass shedding, like in the migration model with an ideal gas
    equation of state, is taken into account. In the inset the baryon mass
    $ M_\mathrm{b} $ versus $ \rho_\mathrm{c} $ relation for this model
    setup is displayed. The models SU (the initial model) and SS (the
    final state for a polytropic equation of state) as well as the
    final state for an ideal gas equation of state are marked. The
    arrows symbolize the respective migration paths.}
  \label{fig:migration}
\end{figure}

Figure~\ref{fig:migration} shows the time evolution of the central
values of the rest-mass density and the lapse. As the star expands,
$ \rho_\mathrm{c} $ decreases while $ N_\mathrm{c} $ grows until the
new stable equilibrium configuration is reached. In the polytropic
case, there are no physical mechanisms to damp the strong pulsations,
and the final state resembles a star oscillating around the
equilibrium configuration until numerical dissipation finally damps
the oscillations. This can be seen in the pulsating values of
rest-mass density and lapse around the value
corresponding to the equilibrium model on the stable branch (solid
horizontal line in Fig.~\ref{fig:migration}).

In the ideal gas case, shock waves are formed at every pulsation, and they
dissipate kinetic energy into thermal energy, thereby damping the
oscillations. As these shocks reach the surface of the star, a small
amount of mass is expelled from the star and matter is ejected
outwards into the surrounding artificial low-density atmosphere until
it leaves the grid across the outer numerical boundary. We
approximately compute the escape velocity as $v_\mathrm{e} =\sqrt{2U}
\approx \sqrt{\psi^2-1}$, where $U$ is the
Newtonian potential. This formula is not exact in general
relativity, but it should by sufficiently accurate near the outer
numerical boundary where gravity is weaker. We find that the
shock waves leaving the computational domain exceed the escape
velocity and therefore the lost mass is gravitationally unbounded. We
also check that these results are not affected by changing the
resolution or setting the outer boundary twice as far away. As the
oscillations are damped, the shocks become weaker and the mass expelled
at each oscillation is smaller. At the end of the simulation the star
has lost about $10\%$ of its initial baryon mass, approaching a state of
constant baryon mass. As a consequence, the final equilibrium configuration
on the stable branch is not the model SS anymore but, rather, the corresponding
model from the stable branch with lower baryon mass and central
density. In Fig.~\ref{fig:migration} we plot the central rest-mass
density and lapse of a series of equilibrium models
on the stable branch corresponding to the baryon mass remaining in the
computational domain at each time. It can be seen that these values
deviate with time from model SS and fit the final state in the
hydrodynamical evolution of the star.

As a by-product of this study we draw reader's attention
to the consistency (as it should be) between the amplitude and the
frequency of the oscillations. The period of these oscillations is
approximately of the order of the hydrodynamical characteristic time
$\tau_\rho$, which decreases with density like $\tau_\rho \approx
\rho^{-1/2}$. In the polytropic case, the maxima of the oscillations
in $\rho_\mathrm{c}$ are systematically higher than in the ideal gas
case. Consequently, the characteristic time is shorter than in the
ideal gas case, as Fig.~\ref{fig:migration} shows. A second property
worth pointing out is that the low numerical viscosity of our
code is responsible for maintaining a nearly constant amplitude of
the oscillations (in the polytropic case) during many characteristic
times.

Our simulations are consistent with the results from a
fully relativistic three-dimensional code in~\cite{Font02}. Similar
simulations of this test, with the original, unmodified CFC scheme, lead 
to a completely incorrect solution with a grossly incorrect
ADM mass. When running with the new improved CFC scheme,
we obtain $M_\mathrm{ADM} = 1.451 M_{\odot}$ and initial values
for the conformal factor and lapse of $\psi_\mathrm{c} = 1.561$ and
$\alpha_\mathrm{c} = 0.273$, respectively. On the other hand, with
the unmodified conventional CFC scheme, the metric solver already
initially converges to a solution with $M_{ADM} = 0.647 M_{\odot}$
(55\%), $\psi_\mathrm{c} = 1.221$ (61\%) and
$\alpha_\mathrm{c} = 0.532$ (63\%), where the relative differences
to the physically correct solution are given in parentheses.

As presented in~\cite{marek_06_a} the migration test can be
successfully simulated using the old CFC scheme, if one resorts to
additionally solving the evolution equation~(\ref{e:der_conf_fact})
for the conformal factor (which would lead to large inconsistencies
in scenarios with higher compactness but still yields acceptable results
for the standard migration case). Here the superiority of the
new, fully consistent CFC scheme, which does not depend on such
scenario-dependend amendments, alredy becomes apparent.


\subsection{Collapse of unstable neutron stars to a black hole}

As the second test we present the collapse of a (spherical or
rotating) neutron star model to a black
hole. Following~\cite{Baiotti05} we trigger the collapse to a black
hole by reducing the polytropic constant $K$ by $2\%$ in the initial
models D1 to D4. Alternatively, in the spherical SU model we increase
the rest-mass density by $0.1 \%$, which yields a similar dynamic
evolution. However, since the models are initially in equilibrium, the
total collapse \emph{time} depends strongly on the perturbation applied. In
these cases, the outer boundary of the finite
difference grid is $20\%$ larger than the star radius. For the
spherical SU model, we perform two simulations using $150$ or $300$
radial cells and $\Delta r_0
\sim 10^{-3}$ or $10^{-4}$, respectively, to assess the resolution
dependence of our simulations. For the rotating models D1 to D4
the grid is made up of $150\times 20$ and $150 \times 40$ cells, with the same
radial grid spacing as in the spherical model. We choose $n_\mathrm{d}=8$ radial domains
for the spectral grid. As in~\cite{Baiotti05} we use a polytropic
equation of state in the evolution.

\begin{figure}[t]
  \epsfxsize = 8.6 cm
  \centerline{\epsfbox{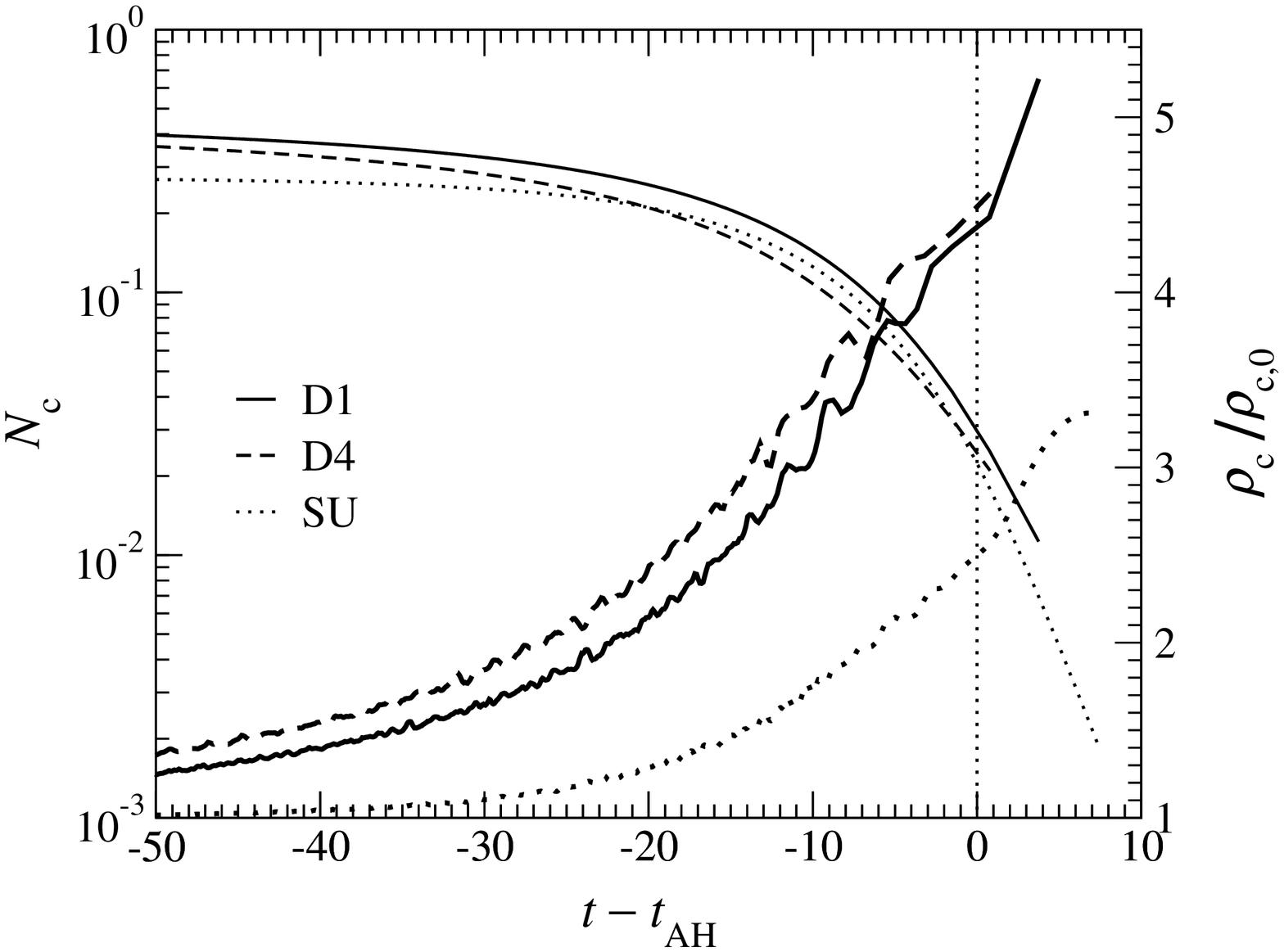}}
  \vspace{1 em}
  \epsfxsize = 8.6 cm
  \centerline{\epsfbox{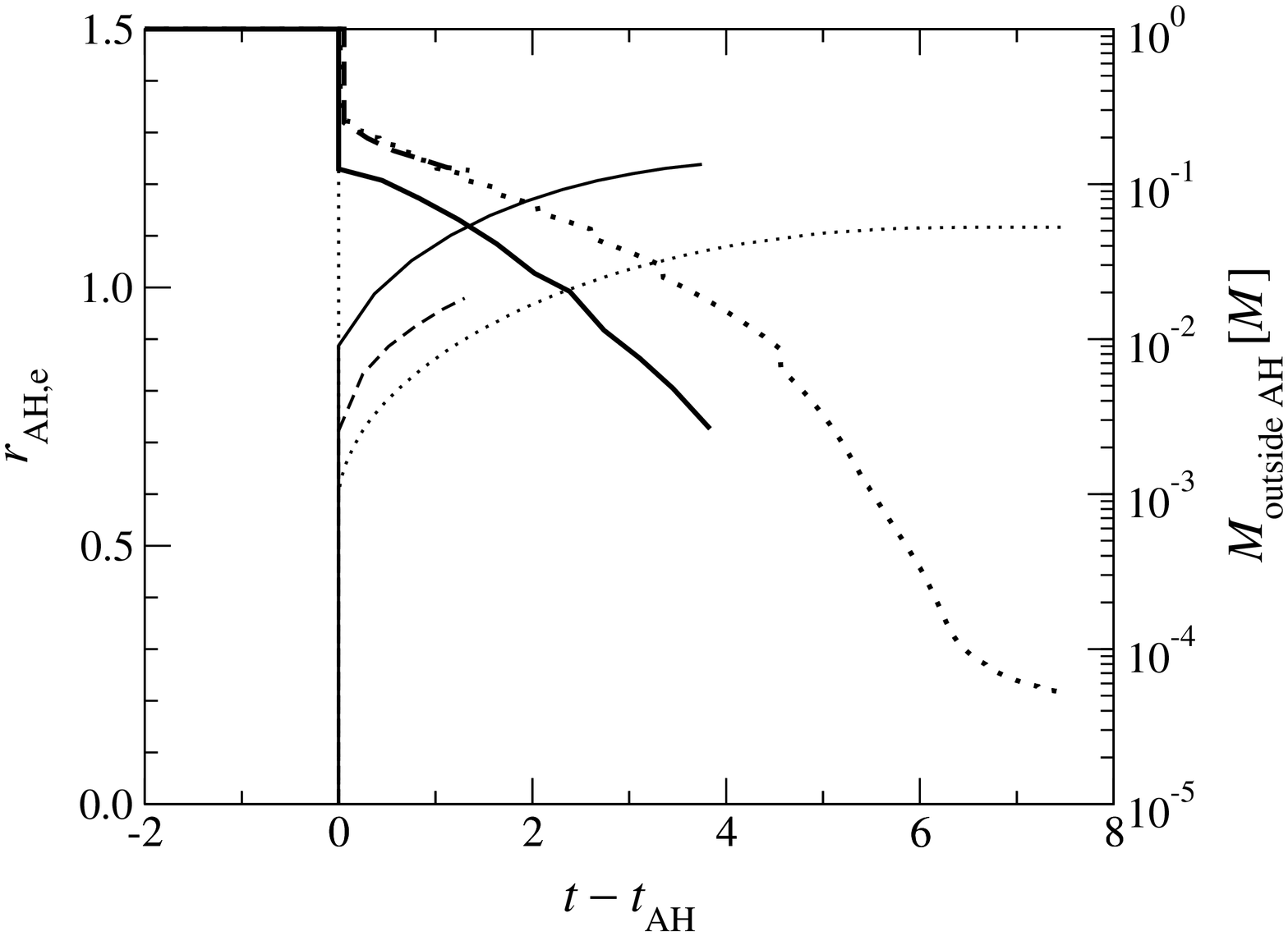}}
  \caption{Collapse to a black hole for the spherical model SU, and the
    rotating models D1 and D4. The top panel shows the time evolution
    of the central lapse $ N_\mathrm{c} $ (thin lines) and the central
    rest-mass density $ \rho_\mathrm{c} $ relative to the initial
    value $ \rho_\mathrm{c,0} $ (thick lines). The bottom panel shows
    the time evolution of the apparent horizon radius
    $ r_\mathrm{AH,e} $ in the equatorial plane (thin lines) and
    the rest mass $ M_\mathrm{outside\,AH} $ remaining outside the
    apparent horizon relative to the total rest mass $ M $ (thick
    lines). The dashed vertical lines mark the time when the apparent
    horizon first appears. If the axes of the lower panel were
    exchanged, the resulting plot would resemble the typical spacetime
    diagram of a star collapsing to a black hole.}
  \label{fig:BH}
\end{figure}

\begin{figure}[t]
  \epsfxsize = 8.6 cm
  \centerline{\epsfbox{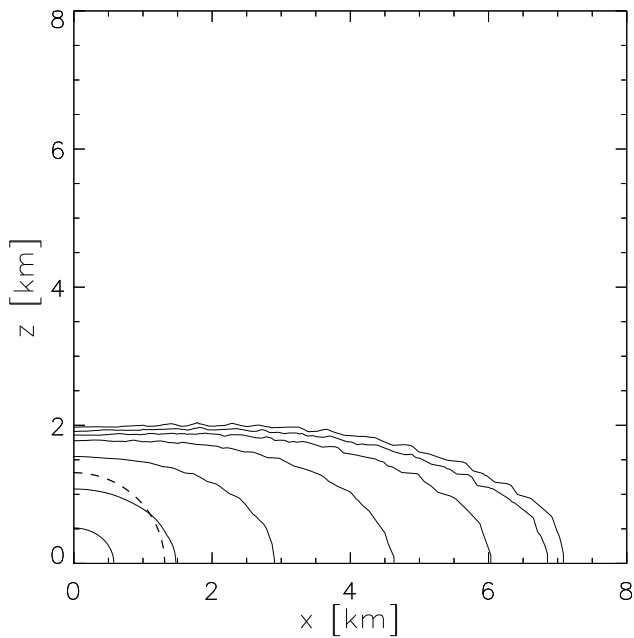}}
  \caption{Isocontours of the rest-mass density for model D4 after the
    apparent horizon first appears at $t=129.9$. The dashed line shows
    the location of the apparent horizon.}
\label{fig:d4}
\end{figure}

The top panel of Fig.~\ref{fig:BH} shows the evolution of the rest-mass
density and lapse at the center. Since for the maximal slicing condition
the singularity cannot be reached in a finite time, $ N_\mathrm{c} $
rapidly approaches zero once the apparent horizon has formed. In
parallel, $ \rho_\mathrm{c} $ grows, which results in a decrease of the
numerical time step due to the Courant condition applied to the
innermost grid cell. We terminate the evolution as the central regions
of the collapsing star inside the apparent horizon become increasingly
badly resolved on the regular grid, and thus numerical errors
grow. We check in SU model that by refining the radial resolution we
are able to follow the collapse to even higher densities. Therefore,
the only limitation to perform a stable evolution after the apparent
horizon formation is the numerical resolution used. Note,
however, that the spatial gauge condition is fixed in CFC, and thus
we are not able to utilize the common method of exploiting the gauge
freedom for the radial component of the shift vector in order to
effectively increase the central resolution.

In the bottom panel of Fig.~\ref{fig:BH} we display the time evolution
of the apparent horizon radius. As expected, the apparent horizon
appears at a finite radius and already encompasses a significant
fraction of the total mass of the star ($\sim 70\mbox{\,--\,}80\%$) at that
time. Afterwards, its radius grows as the surrounding matter falls
inside beyond the horizon. The fraction of the
rest mass remaining outside the horizon is also plotted in the figure.
In the rotating case the apparent horizon is slightly nonspherical.
The ratio of the polar to the equatorial proper circumferential
radius of the apparent horizon at the end of the simulation
is $R_\mathrm{p} /
R_\mathrm{e} = 0.998\mbox{\,--\,}0.978 $ for models D1 to D4, where $R_\mathrm{e} :=
\int_0^{2\pi} \sqrt{g_{\varphi \varphi}} \, d\varphi / (2\pi)$ and $R_\mathrm{p}
:=  \int_0^{\pi} \sqrt{g_{\theta
\theta}} \, d\theta / \pi$.

Since we cannot reasonably determine the location of the event
horizon, as this would require the evolution of spacetime until the
black hole has become practically stationary, we utilize the
apparent horizon radius to estimate the mass
of the newly formed black hole. Following the prescription
in~\cite{Baiotti05} we use the expression $M_\mathrm{BH}=R_\mathrm{e}/2$. Note
that this formula is only strictly valid for a stationary Kerr black
hole. In our case, however, first of all, some (albeit a small) amount
of matter is still outside the horizon and the black hole is still
dynamically evolving, and second, the metric of a Kerr black hole
is not conformally flat~\cite{garat_00_a}. Still, according
to~\cite{Baiotti05} this approximation (excluding the effects of CFC)
introduces an error in the mass estimate of only $\sim 2 \%$.
For the spherical model the estimated value for $ M_\mathrm{BH} $ at the end of the
simulation agrees within $0.5\%$ with the ADM mass $ M_\mathrm{ADM} $ of the
initial model, while in the rotating models D1 to D3 the error is $\le
4\%$. In all these cases the above formula overestimates the black hole 
 mass. Because of its rapid rotation and the
resulting strong centrifugal forces, in model D4 the
collapse deviates significantly from sphericity, leading to a strongly
oblate form of the density stratification. Consequently, we still find
a non-negligible amount of matter outside
the apparent horizon at the end of the simulation (about $12\%$ of the
total rest mass). Therefore the value for $ M_\mathrm{BH} $ is
$8.2\%$ smaller than $ M_\mathrm{ADM} $. In Fig.~\ref{fig:d4} we
present the distribution of the rest-mass density and the location of
the apparent horizon at the end of the simulation for this particular
model. Since the time evolution is limited by our chosen, still
computationally affordable, grid resolution in the central region, we
are not able to evolve this model to times when a disk forms as
in~\cite{Baiotti05}. Nevertheless, all other quantities qualitatively
agree with the results in that work, although we refrain from
performing a more detailed comparison due to the respective
differences in the gauge of the two formulations used
in~\cite{Baiotti05} and in this study, respectively.

In the near future we plan to carry out an exhaustive
analysis of the scenario of a collapse to a black hole by comparing,
on one hand, the CFC formulation with FCF (see
Sec.~\ref{ss:Gen_FCF}), and, on the other hand, the FCF with
other (\emph{free evolution}) formulations. The difficulties induced
by the use of different gauges can be overcome by using
gauge-invariant quantities for comparison and analyzing their
behavior as a function of proper time.


\section{Generalization to the fully constrained formalism}
\label{ss:Gen_FCF}

The ideas presented in Sec.~\ref{sect:schemeCFC} can be generalized
to the FCF approach of the full Einstein equations described in 
Sec.~\ref{ss:brief_review}.

As shown in~\cite{Cordero08},
the hyperbolic part of FCF can be split into a first order system. 
The reformulation of the CFC equations presented in Sec.~\ref{sect:schemeCFC}
relies on the rescaled extrinsic curvature $\hat{A}^{ij}$ given by Eq.~(\ref{e:hatAij}).
Consequently, we write the FCF hyperbolic part as a first order system in $(h^{ij},\hat{A}^{ij})$,
instead of first order system in $(h^{ij},\; \partial h^{ij}/\partial t)$
as in~\cite{Cordero08}, arriving at
\begin{widetext}
  \begin{equationarray}
    \frac{\partial h^{ij}}{\partial t} & = &
    2 N \psi^{-6} \hat{A}^{ij} + \beta^k w^{ij}_k -
    \tilde{\gamma}^{ik} \DSc_k \beta^j -
    \tilde{\gamma}^{kj} \DSc_k \beta^i +
    \frac{2}{3} \tilde{\gamma}^{ij} \DSc_k \beta^k,
    \label{e:dhdt}
    \\
    \frac{\partial \hat{A}^{ij}}{\partial t} & = &
    - \DSc_k \left(-\frac{N \psi^2}{2} \tilde{\gamma}^{kl} w^{ij}_l
    - \beta^k \hat{A}^{ij} \right) -
    \hat{A}^{kj} \DSc_k \beta^i - \hat{A}^{ik} \DSc_k \beta^j +
    \frac{2}{3} \hat{A}^{ij} \DSc_k \beta^k +
    2 N \psi^{-6} \tilde{\gamma}_{kl} \hat{A}^{ik} \hat{A}^{jl}
    \nonumber
    \\ & & - 8 \pi N \psi^6
    \left( \psi^4 S^{ij} - \frac{S \tilde{\gamma}^{ij}}{3} \right) +
    N \left( \psi^2 \tilde{R}_*^{ij} +
    8 \tilde{\gamma}^{ik} \tilde{\gamma}^{jl} \DSc_k \psi \DSc_l \psi \right) + 
    4 \psi \left( \tilde{\gamma}^{ik} \tilde{\gamma}^{jl} \DSc_k \psi \DSc_l N +
    \tilde{\gamma}^{ik} \tilde{\gamma}^{jl} \DSc_k N \DSc_l \psi \right)
    \nonumber
    \\
    & & - \frac{1}{3} \left[ N \left( \psi^2 \tilde{R} +
    8 \tilde{\gamma}^{kl} \DSc_k \psi \DSc_l \psi \right) +
    8 \psi \tilde{\gamma}^{kl} \DSc_k \psi \DSc_l N \right]
    \tilde{\gamma}^{ij}
    \nonumber
    \\
    & & - \frac{1}{2} \left( \tilde{\gamma}^{ik} w^{lj}_k +
    \tilde{\gamma}^{kj} w^{il}_k \right) \DSc_l (N \psi^2) -
    \tilde{\gamma}^{ik} \tilde{\gamma}^{jl} \DSc_k \DSc_l (N \psi^2) +
    \frac{1}{3} \tilde{\gamma}^{ij} \tilde{\gamma}^{kl} \DSc_k \DSc_l (N \psi^2),
    \label{e:dhatAdt}
  \end{equationarray}%
  where
  \begin{equationarray}
    & w^{ij}_k := \DSc_k h^{ij}, &
    \\ [1 em]
    & \tilde{R}_*^{ij} :=
    \frac{1}{2} \left[ - w_l^{ik} w_k^{jl} -
    \tilde{\gamma}_{kl} \tilde{\gamma}^{mn} w_m^{ik} w_n^{jl} +
    \tilde{\gamma}_{nl} w_k^{mn} \left( \tilde{\gamma}^{ik} w_m^{jl} +
    \tilde{\gamma}^{jk} w_m^{il} \right) \right] +
    \frac{1}{4} \tilde{\gamma}^{ik} \tilde{\gamma}^{jl} w_k^{mn}
    \DSc_l \tilde{\gamma}_{mn}. &
  \end{equationarray}%
  The system is closed by adding the equation
  \begin{equation}
    \frac{\partial w^{ij}_k}{\partial t} -
    \DSc_k \left( \beta^l w^{ij}_l + 2 N \psi^{-6} \hat{A}^{ij} \right) =
    - w^{il}_k \DSc_l \beta^j - \tilde{\gamma}^{il} \DSc_k \DSc_l \beta^j -
    w^{lj}_k \DSc_l \beta^i - \tilde{\gamma}^{lj} \DSc_k \DSc_l \beta^i +
    \frac{2}{3} \tilde{\gamma}^{ij} \DSc_k \DSc_l \beta^l +
    \frac{2}{3} w^{ij}_k \DSc_l \beta^l,
    \label{e:dsdtw}
  \end{equation}
  which is derived from applying partial derivatives with respect to $t$ in
  the definition of $w^{ij}_k$. Moreover, the system observes the
  constraint of Dirac gauge, $w^{ij}_i=0$ [Eq.~(\ref{e:gauges})], and for the 
  determinant of the conformal metric, we obtain $\tilde{\gamma}=f$. 
  The first order system given by 
  Eqs.~(\ref{e:dhdt})--(\ref{e:dsdtw}) has the same properties regarding 
  hyperbolicity and existence of 
  fluxes as the one in~\cite{Cordero08}.
  It has the advantage over the second order
  system for $h^{ij}$ proposed in Ref.~\cite{Bonazzola04} of getting
  rid of partial derivatives with
  respect to $t$ of the lapse $N$, the shift $\beta^i$, or the conformal
  factor $\psi$. 
  
  The elliptic part of FCF can be rewritten, using the tensor $\hat{A}^{ij}$, as
  \begin{equationarray}
    \tilde{\gamma}^{kl} \DSc_k \DSc_l \psi & = &
    - 2 \pi \psi^{-1} E^* -  \frac{\tilde{\gamma}_{il}
    \tilde{\gamma}_{jm} \hat{A}^{lm} \hat{A}^{ij}}{8 \psi^{7}} +
    \frac{\psi \tilde{R}}{8},
    \label{e:psiFCF}
    \\
    \tilde{\gamma}^{kl} \DSc_k \DSc_l (N \psi) & = &
    \left[ 2 \pi \psi^{-2} (E^* + 2 S^*) +
    \left( \frac{7 \tilde{\gamma}_{il} \tilde{\gamma}_{jm}
    \hat{A}^{lm} \hat{A}^{ij}}{8 \psi^{8}} +
    \frac{\tilde{R}}{8} \right) \right] (N \psi),
    \label{e:NpsiFCF}
    \\
    \tilde{\gamma}^{kl} \DSc_k \DSc_l \beta^i \! +
    \frac{1}{3} \tilde{\gamma}^{ik} \DSc_k \DSc_l \beta^l & = &
    16 \pi N \psi^{-6} \tilde{\gamma}^{ij}(S^*)_j +
    \hat{A}^{ij} \DSc_j \left( 2 N \psi^{-6} \right) -
    2 N \psi^{-6} \Delta^i_{kl} \hat{A}^{kl}.
    \label{e:shiftFCF}
  \end{equationarray}%
\end{widetext}

The strategy to evolve the two symmetric tensors $h^{ij}$ and
$\hat{A}^{ij}$ relies on a decomposition of these tensors in 
longitudinal and transverse traceless parts. The 
longitudinal parts (divergences with respect to the flat metric) 
are either known a priori or are determined by the elliptic equations.
More specifically, the divergence of  $h^{ij}$ vanishes according to
the Dirac gauge, whereas the divergence of $\hat{A}^{ij}$ is determined
by the momentum constraint~(\ref{e:momentum_hatA}) -- see
below. Consequently, focus is placed on the 
transverse traceless parts of these tensors. The latter are described in a pure-spin
tensor harmonic decomposition, as discussed in a previous
article~\cite{Cordero08}. In particular, each transverse traceless 
tensor is fully expressed in terms of two scalar potentials 
(named $A$ and $\tilde{B}$ in~\cite{Cordero08}) that are evolved
according to evolution equations obtained from the
transverse traceless parts of 
Eqs.~(\ref{e:dhdt}) and~(\ref{e:dhatAdt}) for $h^{ij}$ and
$\hat{A}^{ij}$, respectively, by applying consistently
the decomposition in~\cite{Cordero08}. 
Once the scalar potentials on the next time slice are determined, 
the tensors $h^{ij}$ and ${\hat A}_{\mathrm{TT}}^{ij}$ can be reconstructed
completely, satisfying the divergence-free conditions. This fully
fixes $h^{ij}$, whereas in the case of  
$\hat{A}^{ij}$ the longitudinal part is computed in a very similar
way to the CFC case, i.e.\ by determining the vector $X^i$ from the
momentum constraint as described hereafter.

From Eq.~(\ref{e:hatdefAij}), the momentum constraint can be written as
\begin{equation}
  \DSc_j \hat{A}^{ij} =
  8 \pi \tilde{\gamma}^{ij} (S^*)_j - \Delta^i_{kl} \hat{A}^{kl}
  \label{e:momentum_hatA} , 
\end{equation}
which is equivalent to the following elliptic equation for $X^i$:
\begin{equationarray}
  & & \DSc_j\DSc^j X^i + \frac{1}{3}\DSc^i\DSc_k X^k +
  \tilde{\gamma}^{im} \left( \DSc_k \tilde{\gamma}_{ml} -
  \frac{\DSc_m \tilde{\gamma}_{kl}}{2} \right) \times \quad
  \nonumber
  \\
  & & \qquad \left( \DSc^k X^l + \DSc^l X^k -
  \frac{2}{3} f^{kl} \DSc_p X^p \right) =
  \nonumber
  \\
  & & \quad 8 \pi \tilde{\gamma}^{ij} (S^*)_j - \tilde{\gamma}^{im}
  \left( \DSc_k \tilde{\gamma}_{ml} -
  \frac{\DSc_m \tilde{\gamma}_{kl}}{2} \right)
  \hat{A}_{\mathrm{TT}}^{kl}.
  \label{e:FCF_X}
\end{equationarray}%
This elliptic equation for the vector $X^i$ is linear. Since $h^{ij}$ and
${\hat A}_{\mathrm{TT}}^{ij}$ have been calculated previously, we can solve the
elliptic equation~(\ref{e:FCF_X}) to obtain the vector $X^i$. With this
method, the Dirac gauge and the momentum constraint are guaranteed to be
satisfied. Then, $\hat{A}^{ij}$ is reconstructed from
${\hat A}_{\mathrm{TT}}^{ij}$ and $X^i$ on the new time slice.

At this point, since the tensors $h^{ij}$ and $\hat{A}^{ij}$ are known, 
we can follow exactly the same scheme as in the CFC case to solve in a
hierarchical way the elliptic equations. First the conformal factor is
obtained from Eq.~(\ref{e:psiFCF}), then the lapse function from
Eq.~(\ref{e:NpsiFCF}), and finally the shift vector is acquired from
Eq.~(\ref{e:shiftFCF}). These equations are decoupled
in the order mentioned. 
No sign problems are exhibited in the scalar elliptic equation 
and therefore the maximum principle can be applied. A minor concern
is associated with the sign of the term $\tilde{R}$ in Eq.~(\ref{e:psiFCF}),
but unique solutions also exist for negative conformal Ricci scalars
(closely related to $\tilde{R}$).
Note that, contrary to the CFC case, here no (additional) approximation has
been made: it is simply a new scheme to write down FCF, where the elliptic 
part is better behaved from the point of view of local uniqueness. Numerical
simulations with this FCF scheme will be presented in a future publication.


\section{Discussion}
\label{s:discussion}


\subsection{Summary}

We have presented an approach to the solution of the uniqueness issues
appearing in certain constrained formulations of Einstein equations.
We have illustrated the problem and its solution through a detailed
analytical and numerical study of a waveless approximation that
retains all the involved essential features.

More specifically, we have reformulated XCTS-like elliptic
systems appearing in constrained evolution schemes of the Einstein
equations, like FCF of~\cite{Bonazzola04, Cordero08}, as well as in the
CFC approximation~\cite{Isenberg,WilsoM89}. Such systems require the simultaneous solution of the
constraints, in particular, the momentum constraint for the shift, together
with a maximal slicing condition for the lapse. The resulting elliptic system
presents potential local nonuniqueness problems, and numerical
implementations have indeed encountered such obstacles. The original
CFC formulation has not been able to cope with these
problems, as it suffers from convergence of the system to unphysical solutions
or nonconvergence at all in high density regimes. We have suggested that
these problems are not due to the approximative nature of CFC, since
FCF in the variant of~\cite{Bonazzola04, Cordero08}, which is a
natural generalization of CFC to the nonconformally flat case, 
also suffers from the same problems. In order to address these
issues, first focusing on the simpler CFC case, we
have considered the conformal rescaling of the traceless part of the
extrinsic curvature, resulting in the expression for $\hat{A}^{ij}$ in
Eq.~(\ref{e:hatAij}), which is a rescaling different
from the respective ones employed in FCF and the CFC approximation,
but coincides with the one in the XCTS approach of~\cite{Yor99, PfeYor03}.
This is motivated by the work of Shapiro and
Teukolsky~\cite{Shapiro85}, who simulated the collapse of a neutron
star model using such a reformulation of the CFC metric
equations (however, restricted to spherical symmetry in their case)
and apparently did not encounter any of the problems described above.
Extending their approach to three dimensions, we have decomposed
$\hat{A}^{ij}$ into longitudinal and transverse parts as in the
CTT formulation of the constraint equations~(\ref{e:decomp_tA}). The
divergence (i.e.\ the longitudinal part) of this
tensor is determined by the momentum constraints, Eqs.~(\ref{e:div_hat_A}) in
the CFC case, just as in the CTT formulation. In the CFC scheme, we
have neglected the transverse part of this tensor, as the order of its
error is higher than the one arising from the CFC approximation
itself. In the nonapproximate FCF case, the transverse part
of $\hat{A}^{ij}$ is determined by an evolution equation. Once the
conformal extrinsic curvature is obtained, it can be employed in the
Hamiltonian equation to calculate the conformal factor $\psi$. The lapse is then fixed through the
maximal slicing condition, and the resulting equation allows the application of
a maximum principle uniqueness argument.
Finally, the shift is found through the kinematical relationship
defining the extrinsic curvature, leading to Eq.~(\ref{e:shiftCFC}). 

By performing a variety of tests, we have provided evidence that the
problem of convergence to an unphysical solution of the metric
equations (or even complete nonconvergence) in the original
formulation of the CFC scheme is fully cured by our new
reformulation. Not only can numerical results in the original CFC scheme (in the, at most,
moderately gravitationally compact regime where that system still
yields physically correct solutions) be reproduced by the new
formulation but, more importantly, the new numerical results
presented here exhibit the proper numerical and physical behavior even
for highly compact configurations. For the first time, it has been
possible to successfully perform both the migration test and
the collapse of a neutron star to a black hole in the CFC
case in a consistent way. Our new formulation thus facilitates
simulations in the high density regime of those scenarios where the CFC is
still a reasonably fair approximation, that is, for systems which are
not too far from sphericity, like stellar gravitational collapse.


\subsection{Comparison with previous works}
\label{s:compar_prev_works}

As compared to the original CFC formulation by Isenberg~\cite{Isenberg} and 
Mathews and Wilson~\cite{WilsoM89}, the scheme presented here 
is augmented by an additional vector
elliptic equation for $X^i$, while the elliptic character of the system of
metric equations is preserved. 
The new scheme reformulates the CFC approximation in a CTT shape
(one scalar and one vector elliptic equation), and then solves for the lapse
and the shift (one additional scalar and one vector elliptic equation). In
contrast, the original CFC scheme employed an (X)CTS approach where, together
with two scalar elliptic equations, only one vector elliptic equation was
present. In contrast to the original scheme, 
the elliptic system in the new formulation not only corrects the problem of
local uniqueness in the scalar elliptic equations, but
also introduces a hierarchical structure that decouples the system in
one direction.

In the context of the conformally flat approximation, the same  
``augmented CFC'' scheme as that discussed here has been introduced already
by Saijo~\cite{saijo_04_a} to compute gravitational collapse of differentially rotating
supermassive stars. However, in this work the 
inconsistency between Eq.~(\ref{e:tildeAij}) and Eq.~(\ref{e:decomp_tA}), i.e.\ 
setting to zero the transverse traceless part of $\hat A^{ij}$, has not been pointed out. 
On the contrary, we have analyzed this inconsistency in detail 
(cf.\ the Appendix) and have shown that it leads to an error of 
the same order as that of the CFC approximation. In addition, we have shown here
that the introduction of the vector potential $X^i$ is the key ingredient for solving the nonuniqueness issue. 

The same scheme, but without the conformal rescaling of the matter quantities, has also
been used recently by Shibata 
and Ury\=u~\cite{ShibaU06} in the context of computing initial data. 
As in~\cite{saijo_04_a}, the inconsistency resulting from setting to zero the
transverse traceless part of $\hat A^{ij}$ and the uniqueness issue are
not discussed in their work.
We emphasize that the these studies~\cite{saijo_04_a, ShibaU06} do not discuss the
extension of the new scheme to the nonconformally flat case, as done here. 

Let us also mention that the augmented CFC scheme presented here
can be regarded as a hybrid mixture of some of the waveless
approximation theories (WAT) proposed by Isenberg~\cite{Isenberg}. In fact,
the CFC approximation using the two choices
$\tilde{\gamma}_{ij}=f_{ij}$ and $\partial_t \tilde{\gamma}_{ij}=0$
[as employed in Eq.~(\ref{e:shiftCFC})] corresponds to version
WAT-I. On the other hand, the approximation ${\hat A}_{\mathrm{TT}}^{ij}=0$ used in
Eq.~(\ref{e:decomp_tA}) is in the spirit of the vanishing transverse traceless part of
the extrinsic curvature in the (coupled) version, WAT-II (although
WAT-II refers to the physical extrinsic curvature, whereas here we
have dealt with the conformal one). As mentioned above, both assumptions are
consistent at the considered level of approximation, as shown in
the Appendix.

Regarding the complete constrained evolution of the Einstein
equations, we have generalized the ideas presented here for the
CFC case to the elliptic part of FCF. In previous
studies~\cite{Bonazzola04, Cordero08}, the hyperbolic part
of Einstein equations resulted in a wave-type equation for the tensor
$h^{ij}$, representing the deviation of the three-metric from conformal flatness.
With the introduction of $\hat{A}^{ij}$ we have recovered here a first-order
evolution system, analogous to the standard Hamiltonian $ 3 + 1 $
system, in which we have, however, retained only the divergence-free
terms. Thus, for both $h^{ij}$ and
$\hat{A}^{ij}$, the transverse (divergence-free) parts are evolved by this
system, while the longitudinal parts are fixed either by the gauge (for
$h^{ij}$), or by the momentum constraint (for $\hat{A}^{ij}$). Numerical
results for this case will be presented in future studies.

We finally comment on the recent work by Rinne~\cite{Rinne08}, where
uniqueness problems appearing in certain constrained and partially constrained
schemes for vacuum axisymmetric Einstein equations~\cite{ChoHirLie03b,
  GarDun01} are addressed. 
As in the present case, uniqueness issues related to the Hamiltonian constraint
equation are solved by adopting an appropriate rescaling the extrinsic curvature. On the
other hand, problems associated with the slicing condition are 
tracked to the substitution in that equation of the extrinsic curvature 
by its kinematical expression in terms of the 
(shift and the) lapse. The latter spoils the uniqueness properties by 
reversing the sign of the relevant term in the slicing equation.
This problem is solved by enlarging the elliptic system 
with an additional vector so as to
reexpress the relevant components of
the extrinsic curvature without resorting to the lapse. The 
resulting elliptic system also presents a hierarchical structure.
Although the spirit of such approach is close to the
one here presented, the specific manner of introducing the
additional vector variable in~\cite{Rinne08} critically relies on the
two dimensionality of the axisymmetric problem (specifically, on a choice of a particular
gauge and on the fact that
vectors and rank-two traceless symmetric tensors have
the same number components in two dimensions, a property lost in three dimensions).
On the contrary, the introduction of the vector $X^i$ through the
CTT decomposition~(\ref{e:decomp_tA}) is properly devised to work in three
dimensions. Relevant related discussions in the three-dimensional context
can be found in Sec.~3.4 of~\cite{Rinne08} (where the relation between 
nonuniqueness problems in XCTS and axisymmetric constrained evolution schemes 
is discussed) and in the three-dimensional constrained evolution
scheme presented by Moncrief et~al.\ in~\cite{Moncr07}.


\acknowledgments

It is a pleasure to thank J.~M.~Ib\'a\~nez, M.~Saijo and K.~Ury\=u for many fruitful
discussions. I.~C.-C.\ acknowledges support from from the
Spanish Ministerio de Educaci\'on y Ciencia (AP2005-2857), and
H.~D.\ is supported by the Marie Curie Intra-European Fellowship within the 6th European
Community Framework Programme (IEF 040464). This work was supported by
the Ministerio de Educaci\'on y Ciencia through Grant No.
AYA2007-67626-C03-01, by the Deutsche Forschungsgemeinschaft through
the Transregional Collaborative Research Center Grant No. SFB/TR~7
``Gravitational Wave Astronomy'', by the French ANR Grant No. 06-2-134423
``M\'ethodes math\'ematiques pour la relativit\'e g\'en\'erale'', by
the French--Spanish bilateral research Grant No. HF2005-0115, and by the
European Network of Theoretical Astroparticle Physics ENTApP ILIAS/N6
under Contract No. RII3-CT-2004-506222.


\appendix

\section{Consistency of the approximation}
\label{s:appendix}

In the derivation of the new formalism we make use of the fact that $(L X)^{ij}
\approx \hat{A}^{ij}$ in CFC. We show next that this assumption is
completely consistent at the accuracy level of the CFC approximation. In
the first place, we need to estimate the error of the CFC approximation
itself. By definition, the CFC three-metric deviates linearly with
$h^{ij}$ from the (exact) FCF case. It can be easily shown from the
FCF equations~(\ref{e:psiFCF})-(\ref{e:shiftFCF}) that the metric
quantities behave as
\begin{equationarray}
  \psi & = & \psi_\mathrm{CFC} + \mathcal{O} (h),
  \\
  N & = & N_\mathrm{CFC} + \mathcal{O} (h),
  \\
  \beta^i & = & \beta^i_\mathrm{CFC} + \mathcal{O} (h).
\end{equationarray}%
Therefore $h^{ij}$ can be used as an estimator for the error of the CFC
approximation. 

Two limits in which CFC is exact will be considered. First, in spherical
symmetry the CFC metric system is an exact reformulation of the Einstein
equations since $h^{ij}=0$ in the FCF metric.
If the system is close to spherical symmetry (i.e.\ spheroidal),
and if we are able to define a
quasispherical surface of the system (e.g., the surface of a star or the
apparent horizon of a black hole), then the equatorial and polar circumferential proper
radius, $R_\mathrm{e}$ and $R_\mathrm{p}$, can be computed, and we can
define the ellipticity of the system as
\begin{equation}
  e^2 := 1 - R_\mathrm{p}^2/ R_\mathrm{e}^2.
  \label{e:def_ellipt}
\end{equation}
Close to sphericity, $e^2$ scales linearly with
$h^{ij}$, and we can ensure that the error of CFC is 
$ h^{ij} \sim \mathcal{O}(e^2) $. 
The second limit
to consider is if a post-Newtonian expansion of the gravitational sources is
possible, i.e.\ if the post-Newtonian parameter $\max(v^2/c^2, GM/Lc^2)<1$,
where $v$, $M$, and $L$ are the typical velocity, mass, and length of the system,
respectively. In this case the CFC metric behaves like the first post-Newtonian
approximation~\cite{Kley99, Cerda05}, i.e.\
\begin{equationarray}
  \psi & = & \psi_\mathrm{CFC} + \mathcal{O} \left( 1 / c^4 \right),
  \label{e:pn_psi}
  \\
  N & = & N_\mathrm{CFC} + \mathcal{O} \left( 1 / c^4 \right),
  \label{e:pn_N}
  \\
  c \, \beta^i & = & c \, \beta^i_\mathrm{CFC} + \mathcal{O} \left( 1 / c^4 \right).
  \label{e:pn_betai}
\end{equationarray}%
Note that, for clarity, we explicitly retain powers of the speed of
light $ c $ as factors in the equations throughout this appendix. In
the case that both limits are valid, i.e.\ close to sphericity and in
the post-Newtonian expansion, the
nonconformally-flat part of the three-metric behaves like $h^{ij}\sim
\mathcal{O}(e^2/c^4)$. The next step is to compute the
behavior of the CFC metric if we assume $(LX)^{ij} \approx
\hat{A}^{ij}$, considering the two limiting cases introduced above.

In the spherically symmetric case the relation $(LX)^{ij} =
\hat{A}^{ij}$ is trivially fulfilled. 
Therefore the behavior for
a quasi-spherical configuration is
also $h^{ij}\sim \mathcal{O}(e^2)$ even if ${\hat A}_{\mathrm{TT}}^{ij}=0$
is assumed. This limit in the approximation
is very important, since it is independent of the strength of the gravitational
field. For example, it allows us to evolve black holes, with the only
condition that $h^{ij}$ should be small, i.e.\ close to the sphericity.

To check the approximation in the post-Newtonian limit, we need to
compare $\beta_\mathrm{CFC}^i$ and $X^i$. This can be done by means of the
post-Newtonian expansion of the sources of Eqs.~(\ref{e:beta_cfc0})
and~(\ref{e:equ_X}), respectively,
\begin{equationarray}
  \Delta \beta_\mathrm{CFC}^i + \frac{1}{3} \DSc^i \DSc_j \beta_\mathrm{CFC}^j & = &
  16 \pi S^{*i} + \mathcal{O} \left( 1 / c^7 \right),
  \label{e:pn_beta}
  \\
  \Delta X^i + \frac{1}{3} \DSc^i \DSc_j X^j & = &
  8 \pi S^{*i} + \mathcal{O} \left( 1 / c^7 \right).
  \label{e:pn_X}
\end{equationarray}%
From the comparison of Eqs.~(\ref{e:pn_beta}) and~(\ref{e:pn_X}) we obtain
that
\begin{equation}
  c^3 \, \frac{\beta_\mathrm{CFC}^i}{2} =
  c^3 \, X^i +\mathcal{O} \left( 1 / c^2 \right).
\end{equation}
Thus $\hat{A}^{ij}$ can be computed in terms of $X^i$ as
\begin{equation}
  c^4 \hat{A}^{ij} = \frac{\psi_\mathrm{CFC}^6}{2 N_\mathrm{CFC}} c^4 ( L \beta_\mathrm{CFC})^{ij} =
  c^4 (L X)^{ij} + \mathcal{O} \left( 1 / c^2 \right),
\end{equation}
where we make use of $\psi_\mathrm{CFC}^6/N_\mathrm{CFC} = 
1 + \mathcal{O}(1/c^2)$. The
effect of using $(L X)^{ij}$ instead of $\hat{A}^{ij}$ in the
calculation of the CFC metric can be seen in the expressions
\begin{equationarray}
  \psi_\mathrm{CFC} & = & \Delta^{-1}_\mathrm{s}
  \Sp_{(\psi)} (N_\mathrm{CFC}, \psi_\mathrm{CFC}, \hat{A}^{ij})
  \nonumber
  \\
  & = & \Delta^{-1}_\mathrm{s} 
  \Sp_{(\psi)}(N_\mathrm{CFC}, \psi_\mathrm{CFC}, (LX)^{ij}) 
  \nonumber \\
  &&+ \mathcal{O} \left( 1 / c^8 \right), 
  \label{e:psi}
  \\ [1 em]
  N_\mathrm{CFC} & = & \psi_\mathrm{CFC}^{-1} \Delta^{-1}_\mathrm{s}
  \Sp_{(N\psi)} (N_\mathrm{CFC}, \psi_\mathrm{CFC}, \hat{A}^{ij})
  \nonumber
  \\
  & = & \psi_\mathrm{CFC}^{-1} \Delta^{-1}_\mathrm{s} 
  \Sp_{(N \psi)}(N_\mathrm{CFC}, \psi_\mathrm{CFC}, (L X)^{ij}) 
  \nonumber \\
  &&+ \mathcal{O} \left( 1 / c^8 \right), \qquad \quad
  \label{e:Npsi}
  \\ [1 em]
  c \, \beta_\mathrm{CFC}^i & = & c \, \Delta^{-1}_\mathrm{v}
  \mathcal{S}_{(\beta)} (N_\mathrm{CFC}, \psi_\mathrm{CFC}, \hat{A}^{ij})
  \nonumber
  \\
  & = & c \, \Delta^{-1}_\mathrm{v} 
  \Sp_{(\beta)} (N_\mathrm{CFC}, \psi_\mathrm{CFC}, (L X)^{ij})
  \nonumber \\
  &&+ \mathcal{O} \left( 1 / c^6 \right).
  \label{e:betai}
\end{equationarray}%
where $\Sp_{(\psi)}$, $\Sp_{(N\psi)}$ and
$\Sp_{(\beta)}$ are the sources of
Eqs.~(\ref{e:psiCFC})--(\ref{e:shiftCFC}), and
$\Delta^{-1}_\mathrm{s}$ and $\Delta^{-1}_\mathrm{v}$ are just the
inverse operators appearing in the right-hand-side of these equations
(for the scalars $\psi$ and $N\psi$, and for the vector $\beta^i$,
respectively). When comparing Eqs.~(\ref{e:psi})--(\ref{e:betai})
with Eqs.~(\ref{e:pn_psi})--(\ref{e:pn_betai}), it becomes obvious
that in all cases the error introduced by making the
approximation $(L X)^{ij} \approx \hat{A}^{ij}$ is smaller than the
error of the CFC approximation itself.

\begin{figure}[t]
  \epsfxsize = 8.6 cm
  \centerline{\epsfbox{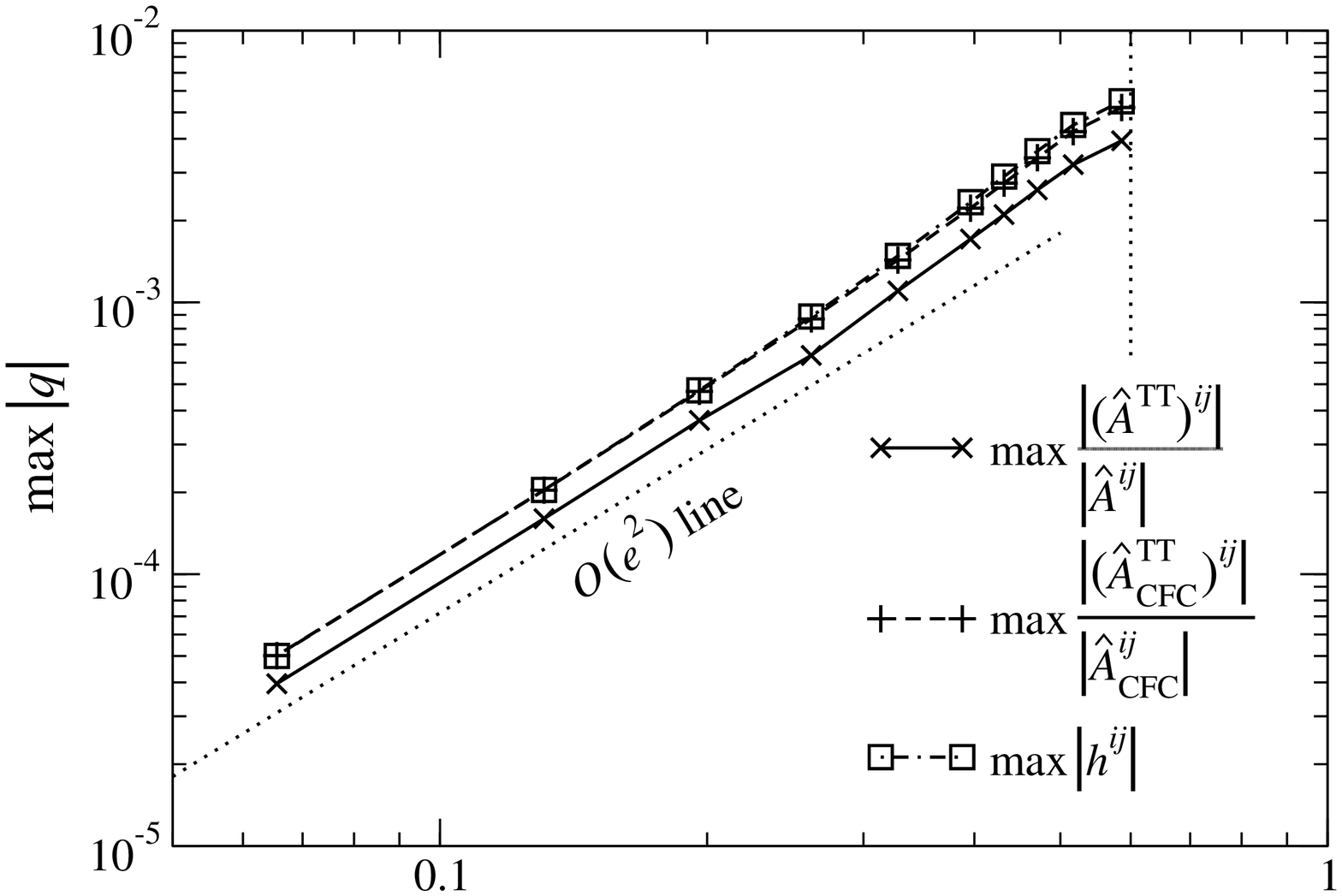}}
  \vspace{1 em}
  \epsfxsize = 8.6 cm
  \centerline{\epsfbox{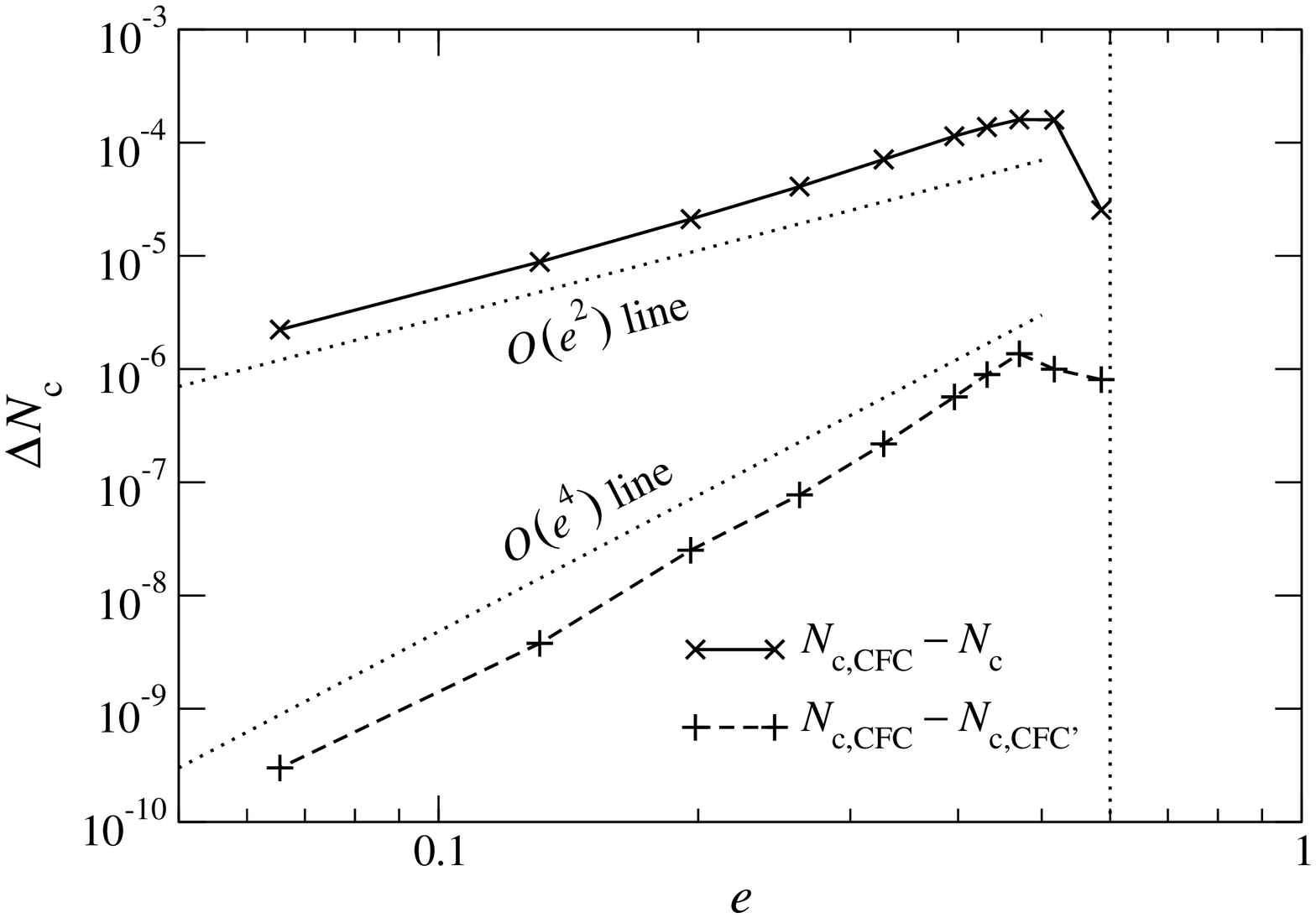}}
  \caption{Consistency of the approximation for rotating neutron star
    models. In the top panel
    $ \max|{\hat A}_{\mathrm{TT}}^{ij}/\hat{A}^{ij}| $ for FCF (solid
    line) and CFC (dashed line) as well as the maximum deviation from
    conformal flatness $ \max|h^{ij}| $ for FCF (dash-dotted line) are
    plotted against the ellipticity $ e $. The bottom panel shows the
    absolute difference $ |N_\mathrm{c,CFC} - N_\mathrm{c}| $
    in the central value of the lapse between CFC and FCF (solid
    line) and the absolute difference
    $ |N_\mathrm{c,CFC} - N_\mathrm{c,CFC'}| $ between regular CFC and
    CFC neglecting ${\hat A}_{\mathrm{TT}}^{ij}$ in
    Eq.~(\ref{e:decomp_tA}) (dashed line). The Kepler limit is marked
    by vertical dotted lines, while the slanted dotted lines represent
    the order of accuracy with respect to powers of $ e $.}
  \label{f:CFC_new_check}
\end{figure}

As an illustration of the above properties, we study the influence of the
${\hat A}_{\mathrm{TT}}^{ij}$ term in Eq.~(\ref{e:decomp_tA}) when computing
rotating neutron star models with a polytropic $\Gamma=2$ equation of state.
This model setup contains the initial models used in
Sec.~\ref{s:numerical_results}. They assume axial symmetry and stationarity,
in combination with rigid rotation. We build a sequence of rotating polytropes
with increasing rotation frequencies, while keeping the central enthalpy
fixed, which produces models of increasing masses from $M=1.33\, M_\odot$ (no
rotation), to $M=1.57 M_\odot$ (the Kepler limit; see below). For all these
models, we use three gravitational field schemes: the exact Einstein equations
using the stationary ansatz in FCF, and the two approximate ones, regular CFC
and CFC, neglecting the term ${\hat A}_{\mathrm{TT}}^{ij}$ in
Eq.~(\ref{e:decomp_tA}). The results are displayed on a logarithmic scale in
Fig.~\ref{f:CFC_new_check}. In the top panel we show the maximal amplitudes of
${\hat A}_{\mathrm{TT}}^{ij}$ (relatively to $\hat{A}^{ij}$) in both FCF and
regular CFC, as functions of the ellipticity $e$ defined in
Eq.~(\ref{e:def_ellipt}). This quantity is physically and numerically limited
by the minimal rotational period at the so-called mass-shedding limit (or
Kepler limit), when centrifugal forces exactly balance gravitational and
pressure forces at the star's equator. In the FCF case we plot the maximal
amplitude of $h^{ij}$. This quantity is dimensionless and represents the
deviation of the three-metric from conformal flatness, which can be interpreted as
the relative error one makes in the metric when using CFC instead of FCF. Note
that this error in computing $\hat{A}^{ij}$ by discarding the
${\hat A}_{\mathrm{TT}}^{ij}$ term in the CFC approximation is roughly of the
same magnitude as the error on the metric in the CFC case. All these
quantities decrease like $\mathcal{O}(e^2)$ as expected, except for stars
rotating close to the Kepler limit. Indeed, the development in powers $e$ is
equivalent to a slow-rotation approximation (see, e.g., \cite{Hartle68}) by
perturbing spherically symmetric configurations, and, when comparing these
slow-rotation results to numerical ``exact'' ones for rigidly rotating stars
(see, e.g., \cite{Prix05} in the two-fluids case), one sees that they usually
agree extremely well, excepted very close to the Kepler limit, where this
``perturbed spherical symmetry'' approach is no longer valid. Finally, because
$\hat{A}^{ij}$ appears as a quadratic source term in the Poisson-like
equations~(\ref{e:psi_cfc0}) and (\ref{e:Npsi_cfc0}), the overall errors on the
lapse $N$ or the conformal factor $\psi$ are even smaller, as shown in the
bottom panel of Fig.~\ref{f:CFC_new_check}. In the case of the central value $
N_\mathrm{c} $ of the lapse, the error due to the CFC approximation is maximal
at the Kepler limit and $\lesssim 10^{-4}$ for the studied sequence. The error
which is then due to neglecting ${\hat A}_{\mathrm{TT}}^{ij}$ within the CFC
scheme amounts to $\lesssim 10^{-6}$ and decreases faster than the error due
to the CFC approximation, namely, as $\mathcal{O}(e^4)$, again except near the
Kepler limit. Our tests thus show that for stationary rotating neutron star
models this additional approximation induces an error which falls within the
overall CFC approximation.


\end{document}